\documentclass[preprint]{elsarticle}
\usepackage[dvips]{color}
\usepackage{subfig}
\newcommand{\bea}{\begin{eqnarray}}
\newcommand{\eea}{\end{eqnarray}}
\newcommand{\be}{\begin{equation}}
\newcommand{\ee}{\end{equation}}
\newcommand{\np}{{\bf p}}

\newcommand{\hk}{\widehat{\bf k}}

\newcommand{\nh}{{\bf h}}
\newcommand{\unit}{{\bf u}}
\newcommand{\nk}{{\bf k}}

\newcommand{\nq}{{\bf q}}

\newcommand{\Qbar}{\not{\!Q}}

\newcommand{\kbar}{\not{\!k}}
\newcommand{\Pbar}{\not{\!P}}
\newcommand{\tauvec}{\mbox{\boldmath $\tau$}}

\newcommand{\Ivec}{\mbox{\boldmath $I$}}

\begin{document}

\title{
Two-nucleon emission 
in neutrino and electron scattering from nuclei:
the modified convolution approximation 
}
\author{I. Ruiz Simo}
\ead{ruizsig@ugr.es}
\author{J.E. Amaro}
\ead{amaro@ugr.es}
\address{Departamento de F\'{\i}sica At\'omica, Molecular y Nuclear,
and Instituto de F\'{\i}sica Te\'orica y Computacional Carlos I,
Universidad de Granada, Granada 18071, Spain}

\author{M.B. Barbaro}
\address{Dipartimento di Fisica, Universit\`a di Torino and
  INFN, Sezione di Torino, \\ Via P. Giuria~1, 10125 Torino, Italy}

\author{J.A. Caballero}
\author{G.D. Megias}
  \address{Departamento de F\'{\i}sica At\'omica, Molecular y Nuclear,
Universidad de Sevilla, Apdo.1065, 41080 Sevilla, Spain}

\author{T.W. Donnelly}
\address{Center for Theoretical Physics, Laboratory for Nuclear
  Science and Department of Physics, Massachusetts Institute of Technology,
  Cambridge, MA 02139, USA}


\date{\today}


\begin{abstract}

The theoretical formalism of inclusive lepton-nucleus scattering in
the two-nucleon emission channel is discussed in the context of a
simplified approach, the modified convolution approximation.  This
allows one to write the 2p2h responses of the relativistic Fermi gas
as a folding integral of two 1p1h responses with the energies and
momenta transferred to each nucleon.  The idea behind this method is
to introduce different average momenta for the two initial nucleons in
the matrix elements of the two-body current, with the innovation that
they depend on the transferred energies and momenta. This method
treats exactly the two-body phase space kinematics, and reduces the
formulae of the response functions from seven-dimensional integrals
over momenta to much simpler three-dimensional ones.  The
applicability of the method is checked by comparing with the full
results within a model of electroweak meson-exchange currents.  The
predictions are accurate enough, especially in the low-energy
threshold region where the average momentum approximation works the
best.

\end{abstract}


\maketitle

\section{Introduction}

The electromagnetic nuclear response for intermediate momentum and
energy transfer is dominated by particle-hole excitations in the
vicinity of the quasielastic peak, located around the energy transfer
$\omega= \sqrt{q^2+m_N^2} -m_N$ needed to knock-out a nucleon
initially at rest with momentum transfer $q$.  But for higher energies
other channels open and the 1p1h description becomes insufficient; in
particular the two-particle two-hole channel starts to play a role as
was first noticed in \cite{Don78,Van81,Alb84}.

In the last decade the increasing interest in the role of multi-nucleon
emission in the electroweak nuclear responses has revealed once more
its importance in describing the kinematical region of the quasielastic
peak and above, and it is at present an active focus of research both
in neutrino and electron scattering studies
\cite{Mar09,Mar10,Nie11,Nie12,Ama11,Ama12}.  In particular, in
charged-current (CC) quasielastic neutrino scattering
$(\nu_\mu,\mu^-)$, the two-particle two-hole (2p2h) channel is now
being considered an essential part in the analysis of the long
baseline experiments \cite{Gra13,Gal16,Sob12,Alv17,Mos16,Kat17}.

In electron scattering the $(e,e'p)$ and $(e,e'pp)$ reactions were
recently measured \cite{Shn07,Sub08,Hen13,Hen14} with the hope of
extracting information on high-momentum components of the reaction
dynamics involving differences between ejection of np and pp pairs of
nucleons.  These experiments have also revitalized interest in
developing models to describe the inclusive 2p2h response function
\cite{Ryc15,Col15,Col16}. More evidence of two nucleon emission of
correlated nucleon pairs has been thought to be found in the ArgoNeuT
neutrino scattering experiment \cite{Acc14}. This has generated
theoretical discussions \cite{Wei16,Nie16}, and it is still under
debate. The most recent theoretical developments of the 2p2h response
functions in neutrino and electron scattering with the shell model has
been reported in \cite{Van16,Van17}.

The first model of 2p2h excitations in the nuclear response can be
traced back to the works of Van Orden et al., \cite{Don78,Van81} who
computed the two-body meson-exchange currents (MEC) contribution in
the non-relativistic Fermi gas model. Later on, Alberico {\it et al.}
used the same model, by adding pionic correlation currents
\cite{Alb84}, obtaining a satisfactory description of the transverse
response functions after including the important enhancement produced
by 2p2h excitations. The first shell model calculations of the
inclusive $(e,e')$ response in the two-nucleon emission channel with
MEC were done by Amaro {\em et al.}  \cite{Ama93,Ama94}.  Several
other improvements including correlation currents, random-phase
approximation and effective interaction were made in
\cite{Alb91,Gil97}.

All of these models were non-relativistic and therefore cannot be applied
to the high energy and momentum transfers of interest for the current
experiments, for which a relativistic description is mandatory.
The first fully relativistic approach to the MEC 2p2h response
function of $^{56}$Fe by Dekker {\em et al.} \cite{Dek91,Dek92,Dek94}
was followed by the Torino model \cite{DePace03,DePace04}, where the
relativistic effects and the scaling properties of the transverse
electromagnetic response were studied. The effect of pionic
correlations was evaluated in \cite{Ama10}.  The validation of the
relativistic MEC model for $(e,e')$ scattering has been recently made
in  \cite{Meg16a}.  These models were extended to the weak sector
in \cite{Rui16} to compute the five CC response functions and the
neutrino inclusive cross section \cite{Meg16b}. In these fully
relativistic models the presence of the $\Delta$
excitation peak without pion emission is evident , which the non-relativistic
models cannot describe in the static limit where the $\Delta$ propagator is constant.

The calculation of the inclusive 2p2h response implies the sum over
all the 2p2h final states.  This involves an integration over all of the
momenta of particles and holes and sums over spin and isospin.  In
general, the complexity of the antisymmetrized two-body current matrix
element prevents the reduction of the dimensionality of the integrals
involved below seven dimensions. But simplifications can be done in
the non-relativistic case, if one neglects the interference terms
between direct and exchange current matrix elements \cite{Van81},
where the integrals are reduced to two dimensions to be performed
numerically.

In the present applications to the neutrino oscillation experiments
the neutrino energy is not fixed and an integral over the neutrino flux
has to be done; this complicates the already cumbersome calculation of
the 2p2h contribution. Therefore, an important goal in such studies is to find simpler
approximations to these response functions in order to reduce the
computational time while keeping the accuracy of the results. 
 This is the motivation of the
present work.

Recently we have developed an approximation which highly simplifies
the calculation of the 2p2h responses, the {\em frozen nucleon
  approximation}. It consists in neglecting the momentum of the
initial nucleons inside the integrals \cite{Rui17}, thus allowing one to
perform analytically a six-dimensional integral over two holes. Assuming
the initial nucleons at rest --- or {\em frozen} --- inside the nucleus
may seem an excessively crude assumption, yet the frozen approximation
works amazingly well for momentum transfers above $q> k_F$, especially
for intermediate and high energy transfer.  This was checked by
comparing with the exact results in a fully relativistic model of
MEC. 

However the frozen approximation fails in the description of the very
low energy transfer region, close to the two-nucleon emission
threshold in the relativistic Fermi gas. For low excitation energy
only the nucleons with momenta close to the Fermi momentum, $k_F$,
contribute. Therefore the frozen assumption is not appropriate in this
energy region. Thus in this work we examine an alternative
procedure, that we have named {\em the modified convolution approximation} 
(MCA), to
describe 2p2h excitations, with good properties in the low energy
region.  It consists in taking an average value for the
momentum of the initial nucleons in the excitation amplitudes, but
treating exactly the kinematics in the phase space.  The average momentum approximation allows one to write the 2p2h response function, namely the imaginary part of the Lindhard function of a nucleon pair, in terms of the Lindhard functions related to each of the two nucleons, which are computed analytically.

Several prescriptions for the the average momentum approximation are
possible. In \cite{Car92}, the photo-absorption cross section in
nuclei was computed by taking the prescription for the average value,
$\langle h \rangle = \sqrt{3/5}k_F$.  In electron scattering the same
prescription was taken in \cite{Gil97} and then for neutrino
scattering in \cite{Nie11}.  This last model includes relativity and
it is considered as benchmark model in the Monte Carlo codes.

However, the average value $\sqrt{3/5}k_F$ is not appropriate for
very low energy transfer, where the momentum of the nucleons is close
to $k_F$.  In the MCA used in this paper we use
a different prescription for the mean value of the initial momenta,
which is compatible with the corresponding energy and momentum
delivered to each one of the two initial nucleons, and therefore it
changes with the kinematics.  Thus we consider that the two nucleons
ejected have averaged momenta $\langle h_1\rangle,
\langle h_2\rangle \ne 0$, for given values of the energy and momentum
transfer $(\omega_1,\nk_1)$ and $(\omega_2,\nk_2)$ to each one of
them, respectively, with
\begin{eqnarray}
\omega &=& \omega_1+\omega_2 \\
q &=& \nk_1+ \nk_2 , 
\end{eqnarray}
where the values of the momenta $\langle h_i \rangle$ depend on
$(\omega_i,\nk_i)$.

The MCA discussed here embodies additional and
interesting features.  First our formalism allows one to include the
exchange diagrams under the average momentum approximation, which is
far from trivial in the formalism of \cite{Nie11,Gil97,Car92}.
Moreover with our formalism we are able to provide for the first time
a test of the average momentum approximation made in
\cite{Nie11,Gil97,Car92} for a wide range of kinematics by comparing
with the exact result using a specific model of MEC. This check was
only made in a particular kinematics for photon absorption in
\cite{Car92}. 

The structure of the work is as follows. In Sect. 2 we review the
formalism of neutrino and electron scattering and the 2p2h response
functions in the relativistic Fermi gas. In Sect. 3 we describe in
detail the MCA. In Sect. 4 we describe the MEC model.  In Sect. 5 we
discuss the treatment of the $\Delta$ propagator.  In Sect. 6 we
compute the 2p2h response functions and compare with the exact
calculation. In Sect. 7 we draw our conclusions.

\section{Formalism of neutrino scattering}
\label{sec_form}

\subsection{Neutrino cross section}

In this work we follow the notations of \cite{Ama05a,Ama05b}.
Here we summarize the formalism for neutrino scattering. The case of
electron scattering can be easily inferred from this by considering
only the relevant longitudinal and transverse response functions.  Thus
we consider charged-current inclusive quasielastic (CCQE) reactions in
nuclei induced by neutrinos and antineutrinos, focusing on the
  $(\nu_\mu,\mu^-)$ and  $(\overline\nu_\mu,\mu^+)$ 
  cross sections.  The relativistic
energies of the incident (anti)neutrino and detected muon are
$\epsilon=E_\nu$, and $\epsilon'=m_\mu+T_\mu$, respectively. Their
momenta are $\nk$ and $\nk'$.  The four-momentum transfer is
$k^\mu-k'{}^\mu=Q^\mu=(\omega,\nq)$, with $Q^2=\omega^2-q^2 < 0$.  The
lepton scattering angle, $\theta$, is the angle between $\nk$ and
$\nk'$.  The double-differential cross section can be written as
\begin{eqnarray}
\frac{d^2\sigma}{dT_\mu d\cos\theta}(E_\nu)
&=&
\left(\frac{M_W^2}{M_W^2-Q^2}\right)^2\frac{G^2\cos^2\theta_c}{4\pi}
\frac{k'}{\epsilon}v_0  
\left[V_{CC} R_{CC}+ \right.
\nonumber\\
&&
\left.
+
2{V}_{CL} R_{CL}+
{V}_{LL} R_{LL}+
{V}_{T} R_{T}
\pm
2{V}_{T'} R_{T'}
\right] \, .
\label{cross}
\end{eqnarray}
Here $G=1.166\times 10^{-11}\quad\rm MeV^{-2}$ is
the Fermi constant, $\theta_c$ is the Cabibbo angle,
$\cos\theta_c=0.975$, and the kinematical factor $v_0=
(\epsilon+\epsilon')^2-q^2$.
The $V_K$ coefficients depend only on the 
lepton kinematics 
and do not depend on the details of the nuclear target:
\begin{eqnarray}
{V}_{CC}
&=&
1+\delta^2\frac{Q^2}{v_0}
\label{vcc}\\
{V}_{CL}
&=&
\frac{\omega}{q}-\frac{\delta^2}{\rho'}
\frac{Q^2}{v_0}
\\
{V}_{LL}
&=&
\frac{\omega^2}{q^2}-
\left(1+\frac{2\omega}{q\rho'}+\rho\delta^2\right)\delta^2
\frac{Q^2}{v_0}
\\
{V}_{T}
&=&
-\frac{Q^2}{v_0}
+\frac{\rho}{2}+
\frac{\delta^2}{\rho'}
\left(\frac{\omega}{q}+\frac12\rho\rho'\delta^2\right)
\frac{Q^2}{v_0}
\\
{V}_{T'}
&=&
-\frac{1}{\rho'}
\left(1-\frac{\omega\rho'}{q}\delta^2\right)
\frac{Q^2}{v_0},
\label{vtp}
\end{eqnarray}
where  we have defined the dimensionless factors
$\delta = m_\mu/\sqrt{|Q^2|}$, proportional to the muon mass $m_\mu$, 
$\rho = |Q^2|/q^2$, and $\rho' = q/(\epsilon+\epsilon')$.

Inside the brackets in Eq. (\ref{cross}) 
there is a linear combination of the five nuclear
response functions, where (+) is for neutrinos and $(-)$ is for
antineutrinos.  The response functions, $R^K(q,\omega)$,
are defined as suitable combinations of the hadronic tensor, $W^{\mu\nu}$, in a
reference frame where the $z$ axis ($\mu = 3$) points along the momentum transfer
$\nq$, and the $x$ axis ($\mu = 1$) is defined as the transverse (to $\nq$) component of
the (anti)neutrino momentum $\nk$ lying in the lepton scattering plane; the $y$ 
axis ($\mu = 2$) is then normal to the lepton scattering plane. The usual components are then
\begin{eqnarray}
R^{CC} &=&  W^{00} \label{rcc} \\
R^{CL} &=& -\frac12\left(W^{03}+ W^{30}\right) \\
R^{LL} &=& W^{33}  \\
R^{T} &=& W^{11}+ W^{22} \\
R^{T'} &=& -\frac{i}{2}\left(W^{12}- W^{21}\right)\,. \label{rtprima}
\end{eqnarray}
The hadronic tensor is the core of our calculation.

\subsection{Hadronic tensor}

The inclusive hadronic tensor is constructed from bilinear combinations of matrix elements of 
the current operator $J^\mu(Q)$, summing over
all the possible final nuclear states with excitation energy
$\omega=E_f-E_i$ 
\begin{equation}
W^{\mu\nu}
= 
\sum_f \overline{\sum_i} 
\langle f | J^\mu(Q) |i \rangle^*
\langle f | J^\nu(Q) |i \rangle
\delta(E_i+\omega-E_f) .
\end{equation}

In this work we consider the (non-interacting) relativistic Fermi gas
model of the nucleus. The nuclear states are Slater determinants
constructed with single-particle (Dirac) plane waves states. All states
with momentum $h<k_F$ are occupied in the ground state.  Within this
model the final nuclear states can be one-particle one-hole
(1p1h), 2p2h, and so on. Therefore the hadronic tensor can be
expanded as
\begin{equation} 
W^{\mu\nu}
= 
W^{\mu\nu}_{\rm 1p1h}
+ W^{\mu\nu}_{\rm 2p2h} + \cdots
\end{equation}
The simplest excited states, 
1p1h, are constructed by raising a particle above the Fermi level, with momentum
$p'>k_F$, leaving a hole with momentum $h<k_F$.  These final states
contribute to the typical quasielastic peak shape of the hadronic tensor 
$W^{\mu\nu}_{\rm 1p1h}$ in the impulse approximation, where the current
$J^\mu(Q)$ is a one-body operator.

In this work we focus on the 2p2h part of the hadronic tensor, which
contributes to two-nucleon emission. To get this we need a two-body
current operator, whose matrix elements are given by
\begin{eqnarray}
&& 
\kern -2cm 
\langle P'_1P'_2 | J^\mu(Q) | H_1H_2\rangle =
\nonumber\\
&&
\kern -1cm 
\frac{(2\pi)^3}{V^2}\delta(\np'_1+\np'_2-\nq-\nh_1-\nh_2)
\frac{m_N^2}{\sqrt{E'_1E'_2E_1E_2}}
j^{\mu}(\np'_1,\np'_2,\nh_1,\nh_2),
\end{eqnarray}
where $V$ is the volume of the system and we have defined the four-vectors 
of the particles and holes, as $P'_i=(E'_i,\np'_i)$, and
$H_i=(E_i,\nh_i)$, respectively, for $i=1,2$.  Note that the above
matrix element conserves three-momentum because our wave functions are
plane waves. The relativistic boost factors $(m_N/E)^{1/2}$ are
factorized out of the spin-isospin dependent two-body current
functions $j^{\mu}(\np'_1,\np'_2,\nh_1,\nh_2)$, which are defined by
the above expression.

Inserting this expression into the definition of the hadronic tensor
for 2p2h final states and taking the thermodynamic limit
$V\rightarrow \infty$, we obtain
 \begin{eqnarray}
W^{\mu\nu}_{\rm 2p2h}
&& 
=\frac{V}{(2\pi)^9}\int
d^3p'_1
d^3p'_2
d^3h_1
d^3h_2
\frac{m_N^4}{E_1E_2E'_1E'_2}
\nonumber \\ 
&&
\times w^{\mu\nu}(\np'_1,\np'_2,\nh_1,\nh_2)\;
\delta(E'_1+E'_2-E_1-E_2-\omega)
\nonumber\\
&&
\times 
\Theta(p'_1,h_1)\Theta(p'_2,h_2)
\delta(\np'_1+\np'_2-\nq-\nh_1-\nh_2) \, ,
\label{hadronic12}
\end{eqnarray}
where $V/(2\pi)^3 \frac8 3 \pi k_F^3 = Z$ for symmetric nuclear matter
with the Fermi momentum  $k_F$. 
Here we have defined the Pauli blocking function $\Theta$ as 
the product of step-functions 
\begin{equation}
\kern -8mm
\Theta(p',h) \equiv
\theta(p'-k_F)
\theta(k_F-h) .
\end{equation}

The function $w^{\mu\nu}(\np'_1,\np'_2,\nh_1,\nh_2)$ represents the
hadron tensor for the elementary 2p2h transition of a nucleon pair
with given initial and final momenta, summed up over spin and
isospin,
\begin{eqnarray}
\kern -8mm
w^{\mu\nu}(\np'_1,\np'_2,\nh_1,\nh_2) &=& \frac{1}{4}
\sum_{s_1s_2s'_1s'_2}
\sum_{t_1t_2t'_1t'_2}
j^{\mu}(1',2',1,2)^*_A
j^{\nu}(1',2',1,2)_A \, , \nonumber \\
&&
\label{elementary}
\end{eqnarray}
which is written in terms of the antisymmetrized two-body current
matrix elements 
\begin{equation} \label{anti}
j^{\mu}(1',2',1,2)_A
\equiv j^{\mu}(1',2',1,2)-
j^{\mu}(1',2',2,1) \,.
\end{equation}
 The factor $1/4$ in Eq.~(\ref{elementary}) accounts for the
 antisymmetry of the two-body wave function.  Note that the exchange
 $1\leftrightarrow 2$ in the second term implies implicitly the
 exchange of momenta, spin and isospin quantum numbers.

To compute the inclusive 2p2h response functions
we integrate over $\np'_2$ using the momentum delta-function, 
finally obtaining
\begin{eqnarray}
R^{K}_{\rm 2p2h}
&=&
\frac{V}{(2\pi)^9}\int
d^3p'_1
d^3h_1
d^3h_2
\frac{m_N^4}{E_1E_2E'_1E'_2}
\Theta(p'_1,h_1)\Theta(p'_2,h_2)
\nonumber \\ 
&&
\times r^{K}(\np'_1,\np'_2,\nh_1,\nh_2)\;
\delta(E'_1+E'_2-E_1-E_2-\omega) ,
\label{hadronic}
\end{eqnarray}
where $\bf p'_2= h_1+h_2+q-p'_1$ by momentum conservation.  The five
elementary response functions for a 2p2h excitation $r^K$ are defined
in terms of the elementary hadronic tensor $w^{\mu\nu}$ as in Eqs.
(\ref{rcc}--\ref{rtprima}), for $K=CC,CL,LL,T,T'$.  The five inclusive
responses embody a global axial symmetry around the $z$ axis defined
by $\nq$. This allows us to fix the azimuthal angle of one of the particles.
We choose to integrate over the angle of the  particle $\np'_1$
setting 
$\phi'_1=0$. 
Consequently the integral over $\phi'_1$
gives a factor $2\pi$. Furthermore, the energy delta-function enables
analytical integration over $p'_1$, and so the integral in
Eq.~(\ref{hadronic}) can be reduced to seven dimensions. In the
``exact'' results shown in the next section, this 7D integral has been
computed numerically using the method described in
\cite{Sim14}.

An approximation was made in \cite{Rui17}, consisting in setting 
$\nh_1=\nh_2=0$, and $E_1=E_2=m_N$ and thereby allowing one to perform the integral
over $\nh_1,\nh_2$.  This limit corresponds to the frozen nucleon
approximation, which does not properly describe the threshold region
for small values of $\omega$.  The frozen response functions $R_{\rm
  frozen}^K$, are given by
\begin{eqnarray}
\kern -4mm 
R^{K}_{\rm frozen}
&=& \frac{V}{(2\pi)^9}
\left(\frac43\pi k_F^3\right)^2
\int d^3p'_1\;
\frac{m_N^2}{E'_1E'_2}\;
r^{K}(\np'_1,\np'_2,0,0)
\nonumber\\
&\times&
\delta(E'_1+E'_2-2m_N-\omega)\;
\Theta(p'_1,0)\Theta(p'_2,0) .
\label{tensor_frozen}
\end{eqnarray}
This integral can be reduced to 
one dimension, which is convenient for applications to neutrino scattering, at least at high $\omega$. 

In this work we are interested in improving this frozen approximation,
by treating exactly the phase space dependence implied by the energy-conserving delta-function, which in the frozen approximation neglects
the motion of the initial nucleons, and therefore modifies the
argument of the delta-function.

\section{The modified convolution approximation}

In the MCA we introduce an average momentum for
the initial nucleons, but we treat exactly the energy balance between
the particle and hole momenta, contrary to the frozen
approximation where the initial momenta are approximated by zero also in
the kinematics.

The procedure consists in splitting the energy delta-function into an
integral of two delta-functions over the energy transfer $\omega_1$ to
the first nucleon:
\begin{equation}
\delta(E'_1+E'_2-E_1-E_2-\omega)
= \int_0^\omega
d\omega_1 
\delta(E'_1-E_1-\omega_1)
\delta(\omega_1+E'_2-E_2-\omega) .
\end{equation}
Inserting this relation in Eq. (\ref{hadronic})
we can write the response function 
as
 \begin{eqnarray}
R^{K}_{\rm 2p2h}
&=&
\int
d^3p'_1
d^3h_1
d^3h_2
\Theta(p'_1,h_1)\Theta(p'_2,h_2)
\nonumber \\ 
&\times&
\int_0^\omega
d\omega_1 
\delta(E'_1-E_1-\omega_1)
\delta(\omega_1+E'_2-E_2-\omega)
f^K ,
\label{cambio1}
\end{eqnarray}
where we use the short notation for the integrand containing the elementary 
hadronic tensor and phase space factors
\begin{equation}
f^K \equiv \frac{V}{(2\pi)^9}
\frac{m_N^4}{E_1E_2E'_1E'_2}
r^{K}(\np'_1,\np'_2,\nh_1,\nh_2) .
\end{equation}
Now for $\nh_1$ fixed we change the variable $\np'_1 \rightarrow
\nk_1$, the momentum transfer to the first nucleon. Thus
\begin{eqnarray}
\np'_1 = \nh_1+\nk_1   \\
\np'_2 = \nh_2+\nq-\nk_1 .
\end{eqnarray}
Performing this change in Eq. (\ref{cambio1}) and reordering the
integrations we obtain
 \begin{eqnarray}
R^{K}_{\rm 2p2h}
&=&
\int
d^3k_1
\int_0^\omega
d\omega_1 
\int
d^3h_1
\Theta(|\nh_1+\nk_1|,h_1)
\delta(E'_1-E_1-\omega_1)
\nonumber \\ 
&&
\int
d^3h_2
\Theta(|\nh_2+\nq-\nk_1|,h_2)
\delta(\omega_1+E'_2-E_2-\omega)
f^K .
\end{eqnarray}
Now we assume that the elementary pair response function $r^K$ can be
approximated by its value at some average momenta $\langle \nh_1
\rangle$ and $\langle \nh_2 \rangle$ to be specified below:
\begin{equation}
\langle r^{K} \rangle \equiv
r^{K}(\np'_1,\np'_2,\langle\nh_1\rangle,\langle\nh_2\rangle)\; ,
\end{equation}
where now
\begin{equation}
\np'_1 = \langle\nh_1\rangle+\nk_1,
\qquad
\np'_2 = \langle\nh_2\rangle+\nq-\nk_1.
\end{equation}
Then the integrals over $\nh_1$ and $\nh_2$ can be performed
separately, yielding
 \begin{eqnarray}
R^{K}_{\rm 2p2h}
&=&
 \frac{V}{(2\pi)^9}m_N^4
\int
d^3k_1
d\omega_1 
\langle r^{K}\rangle 
R_{\rm 1p1h}(k_1,\omega_1) R_{\rm 1p1h}(k_2,\omega_2) \, ,
\label{halfrozen}
\end{eqnarray}
where we have defined 
$\nk_2$ and $\omega_2$ as the momentum 
and energy transferred to the second nucleon
\begin{equation}
\nk_2 = \nq-\nk_1,
\qquad
\omega_2= \omega-\omega_1 
\end{equation}
and the (dimensionless) elementary 1p1h response function is given by
\begin{equation}
R_{\rm 1p1h}(q,\omega) =
\int d^3h
\Theta(|\nh+\nq|,h)
\frac{1}{E_hE_{|\nh+\nq|}}
\delta(E_{|\nh+\nq|}-E_h-\omega) .
\label{r1p1h}
\end{equation}
Note that this elementary response function is proportional to the
imaginary part of the relativistic Lindhard function, for symmetric
matter, which can be found in Appendix B of  \cite{Nie04}:
\begin{equation}
R_{\rm 1p1h}= -\left(\frac{2\pi}{m_N}\right)^2 {\rm Im} \overline{U}_R(q,\omega) .
\end{equation}
Equation (\ref{halfrozen}) corresponds to the MCA of
the 2p2h inclusive responses. They are written as a convolution of
the elementary 1p1h responses for single excitation of each nucleon,
modified by a weight function.
This represents the average excitation response of the pair, 
$\langle r^{K}\rangle$. Each 1p1h response carries  
the correct energy and momentum applied to each nucleon, globally
sharing the total energy and momentum transferred to the pair,
$(q,\omega)$.

This formula allows one to relate directly the 2p2h model of \cite{Rui16}
with the diagrammatic formalism of  \cite{Nie11,Gil97,Car92},
which in an alternative way also factorizes the separate Lindhard
functions for the direct many-body diagrams by using the Cutkosky
rules.  Thus it will be useful to compare results from these two
different formalisms. Note that Eq. (\ref{halfrozen}) also includes the
exchange diagrams, which are implicit in the pair elementary responses.

In summary, the prescription of \cite{Nie11} was to use a constant average
value $\langle h_1 \rangle = \langle h_2 \rangle = (3/5)^{1/2} k_F$.
On the contrary, in this approach we specify a different prescription,
with different average values for $h_1$ and $h_2$, that depend on the
kinematics, as explained below.

\subsection{Value of the averaged momentum}

To obtain a proper value of the average hole momentum it is convenient
briefly to recall the essential points of the analytical integration of
the 1p1h response function in a relativistic Fermi gas.

We start from Eq. (\ref{r1p1h}) by changing variables 
$(h,\theta,\phi) \rightarrow (E,E',\phi)$, where 
\begin{eqnarray}
E^2 & = &  h^2 + m_N^2 \\
E'{}^2 & = & (\nh+\nq)^2 + m_N^2 = m_N^2+h^2+q^2+2hq\cos\theta .
\end{eqnarray}
The volume element becomes
\begin{equation}
d^3h = \frac{EE'}{q}dE dE' d\phi .
\end{equation}
Then the integral over $\phi$ gives $2\pi$ and the response function is
\begin{equation}
R_{\rm 1p1h}= \frac{2\pi}{q}\int_{m_N}^{E_F}dE
\int_{E_{h-q}}^{E_{h+q}}dE'
\delta(E'-E-\omega)\theta(E'-E_F) \, ,
\end{equation}
where $E_F$ is the relativistic Fermi energy.
Integrating the $\delta$-function one has $E'=E+\omega$ and 
\begin{equation}
R_{\rm 1p1h}= \frac{2\pi}{q}\int_{m_N}^{E_F}dE
\theta(E_{h+q}-E-\omega)\theta(E+\omega-E_{h-q})
\theta(E+\omega-E_F) .
\end{equation}
The first two step-functions inside the integral
imply  the following inequalities
\begin{equation}
E_{h-q} < \omega + E < E_{h+q} \, ,
\end{equation}
which is just a consequence of energy-momentum conservation. This can be
shown to be equivalent to the single condition
\begin{equation}
\kappa \sqrt{1+1/\tau } - \lambda < \epsilon \, ,
\end{equation}
where for convenience we use dimensionless variables defined by
\begin{equation} \label{adimensional}
\epsilon
 = \frac{E}{m_N}, \quad
\kappa = \frac{q}{2m_N} , \quad
\lambda = \frac{\omega}{2m_N} , \quad
\tau = \kappa^2 - \lambda^2.
\end{equation}
On the other hand, the last step-function inside the integral implies that
\begin{equation}
\epsilon_F - 2\lambda < \epsilon \, ,
\end{equation}
where $\epsilon_F = E_F/m_N$ is the Fermi energy in units of the
nucleon mass. Performing the change of variable $E \rightarrow
\epsilon = E/m_N$, the above integral can be written as
\begin{equation}
R_{\rm 1p1h}= \frac{2\pi}{q}m_N\int_{\epsilon_0}^{\epsilon_F} d\epsilon
\theta(\epsilon_F-\epsilon_0) =
\frac{\pi}{\kappa}(\epsilon_F-\epsilon_0)\theta(\epsilon_F-\epsilon_0) \, ,
\label{r1p1h-integrada}
\end{equation}
where we have defined the lower limit as
\begin{equation} \label{epsilon0}
\epsilon_0 = {\rm Max}\left\{ \kappa \sqrt{1+1/\tau } - \lambda ,
\epsilon_F-2\lambda\right\} .
\end{equation}
From Eq. (\ref{r1p1h-integrada})  
it is evident that the initial energy of the nucleon is
restricted to fall between the limits
\begin{equation}
\epsilon_0 m_N < E < \epsilon_F m_N .
\end{equation}
The mean value of the energy in this interval  is
\begin{equation}  \label{mean-energy}
\langle E \rangle 
= \frac{\epsilon_0+\epsilon_F}{2}m_N \, ,
\end{equation}
and this provides our choice 
for the average hole momentum in the MCA
\begin{equation} \label{mean-momentum}
\langle h \rangle ^2 = \langle E \rangle ^2 - m_N^2 \, .
\end{equation}
It is also convenient to write the 1p1h response function in terms of
the scaling variable defined by
\begin{equation} \label{psi2}
\psi^2 = \frac{\epsilon_0-1}{\epsilon_F-1} < 1
\end{equation}
or equivalently
\begin{equation}
\epsilon_0 = 1 +\psi^2\xi_F \, ,
\end{equation}
where $\xi_F= \epsilon_F-1$ is the Fermi kinetic energy in units of
nucleon mass.  Then we obtain
\begin{equation}
R_{\rm 1p1h} 
= \frac{\pi}{\kappa}\xi_F
(1-\psi^2)\theta(1-\psi^2) .
\label{r1p1h-scaling}
\end{equation}
As a function of $\psi^2$ the response function is an inverted
parabola in the region $-1 < \psi < 1$ and it is zero outside this interval. 
The maximum corresponds to the
center of the quasielastic peak, for $\psi=0$ or $\epsilon_0=1$. This
implies that the momentum of the nucleon at the QE peak can take on 
all the values between zero and $k_F$. When we depart from the center and
approach the borders defined by $\psi=\pm 1$, the value of $\epsilon_0$
approaches $ \epsilon_F$, and therefore the value of the momentum
$h$ of the hole is more restricted below $k_F$. At the borders it is
exactly $k_F$. Therefore there exists a region of $\omega$ values close
to the borders of the QE peak where the momentum of the hole is always
larger than $(3/5)^{1/2}k_F$, which is the average value employed in  \cite{Nie11,Gil97,Car92}.

Following the average momentum definition,
Eqs. (\ref{mean-energy},\ref{mean-momentum}), in the MCA we
compute two different momenta, $\langle h_1 \rangle$ and $\langle
h_2\rangle$, depending on  the momentum and energy transfer to each
nucleon, $k_i,\omega_i$. To compute $\langle h_i \rangle$,
we must evaluate the minimum nucleon energy  in Eq.~(\ref{epsilon0}) 
using the dimensionless variables in Eq.~(\ref{adimensional}) for
$q=k_i$ and $\omega=\omega_i$.

\subsection{Direction of the averaged momentum}

In this subsection, for simplicity, 
we use the notation $\nh_i$ for the averaged hole momenta. 
The above discussion allows us to 
determine the modulus of the averaged momenta
$ h_i$. Concerning its direction, it is only
 possible to determine
the angle between $\nh_i$ and $\nk_i$.   By imposing 
energy conservation for an on-shell nucleon with initial momentum $\nh_i$, 
\begin{equation}
E'_i=E_i+\omega_i = \sqrt{m_N^2+(\nh_i+\nk_i)^2} ,
\end{equation}
and, taking the square
\begin{equation}
(E_i+\omega_i)^2= m_N^2+h_i^2+k_i^2+2h_ik_i\cos\theta_i ,
\end{equation}
we get the angle between $\nh_i$ and $\nk_i$:
\begin{equation}\label{coseno}
\cos\theta_i= \frac{\omega_i^2+2E_i\omega_i-k_i^2}{2h_ik_i} .
\end{equation}
By construction of the MCA average momentum $\langle h_i\rangle$
given by Eqs. (\ref{mean-energy},\ref{mean-momentum}), the above value
of the angle is within the correct limits $-1\le \cos\theta_i \le 1$.
Note that using a constant average momentum such as $\sqrt{3/5}k_F$
there are kinematics in $(k_i,\omega_i)$ where the above angle is
undefined because it is outside the region allowed by energy conservation.

\begin{figure}[t]
\centering
\includegraphics[width=9cm,bb=230 560 380 680]{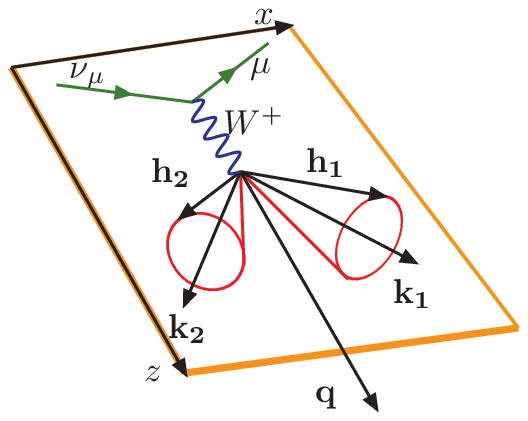}
\caption{ Geometry defining the lepton scattering plane and the two
  cones spanning the possible momenta of the two holes around the
  momenta transferred to each nucleon, $\nk_i$. The cones are
  determined by the energy transferred $\omega_i$ to each nucleon and
  by the average hole momenta, $h_i$. The final momenta $\np'_i=\nh_i+\nk_i$
  are not shown for simplicity.  }\label{conos}
\end{figure}

For the total determination of the averaged momenta, $\nh_i$, one should know 
in addition the azimuthal angles with respect to the $\nk_i$ vectors.
But from the 1p1h response function $R_{\rm 1p1h}(k_i,\omega_i)$ there are not
restrictions over the azimuthal angles. 
Energy-momentum conservation only
provides restrictions for the magnitude of the initial momenta and
their angles with respect to $\nk_i$.
For a given angle $\theta_i$ between $\nh_i$ and $\nk_i$, Eq.~(\ref{coseno}), the vector form of  $\nh_i$ is
\begin{equation}
\nh_i= h_i (\cos\theta_i \hk_i + \sin\theta_i \unit_i) ,
\end{equation}
where $\hk_i$ is the unit vector in the direction of $\nk_i$ and $\unit_i$
 is an unit vector perpendicular to $\nk_i$. 
The vectors $\nh_i$ generate two cones around the $\nk_i$ vectors, as depicted in Fig.  \ref{conos}.
In our reference frame (see below) the vector $\nk_i$ can be considered 
in the scattering plane, spanned by the $x,z$ directions, 
as shown in Fig. \ref{conos}, given by
\begin{equation} 
\nk_i= (k_i^x,0,k_i^z) .
\end{equation}
Therefore the general form of the unit vector in the plane perpendicular to $\nk_i$ is
\begin{equation} \label{ui}
\unit_i = 
\pm \frac{(-k_i^z,\alpha_i,k_i^x)}{\sqrt{(k_i^z)^2+\alpha_i^2+(k_i^x)^2}} \, ,
\end{equation}
where $\alpha_i$ is a real parameter that determines the $y$ component
of $\nh_i$.

There are no restrictions over the values of the two parameters
$\alpha_1$, $\alpha_2$.  
 In practice what we do is to choose several
options for these parameters guided by simplicity of the
calculation. The simplest option is to choose $\alpha_1=\alpha_2=0$,
but any other election is possible.  In the results section we compare
several options and discuss which is the best one according to the
comparison with the full results, and study how the results depend on
the values of $\alpha_i$.

\subsection{MCA Integration limits}

To evaluate the MCA expression for the 2p2h responses, Eq. (\ref{halfrozen}), 
it is convenient to change the integration variable $\theta_{k_1}$
(the angle between $\nk_1$ and $\nq$) to the magnitude, $k_2$, of the 
momentum transfer to the second nucleon
\begin{equation}
\cos\theta_{k_1} \longrightarrow k_2 = |\nq-\nk_1| .
\end{equation}
The Jacobian of the transformation gives
\begin{equation}
k_1^2 dk_1 d\cos\theta_{k_1}d\phi_{k_1} = 
\frac{k_1k_2}{q}dk_1dk_2d\phi_{k_1}.
\end{equation}
Due to the azimuthal symmetry of the inclusive response functions, the
integration over the  angle $\phi_{k_1}$ can be reduced 
 to multiplication by $2\pi$ and evaluation of the integrand 
for $\phi_{k_1}=0$, as a particular case. 
The MCA reduces to a three-dimensional integral given by
 \begin{eqnarray}
R^{K}_{\rm 2p2h}
&=&
 \frac{V}{(2\pi)^9}
\frac{2\pi m_N^4}{q}
\int^{(k_1)_{\rm max}}_0
dk_1 k_1
\int^{q+k_1}_{|q-k_1|}
dk_2 k_2
\nonumber\\
&& 
\int^{(\omega_1)_{\rm max}}_{(\omega_1)_{\rm min}}
d\omega_1 
\langle r^{K}\rangle 
R_{\rm 1p1h}(k_1,\omega_1) R_{\rm 1p1h}(k_2,\omega-\omega_1) .
\label{MCA}
\end{eqnarray}

\begin{enumerate}

\begin{figure}
\centering
\includegraphics[width=12cm,bb=60 250 550 790]{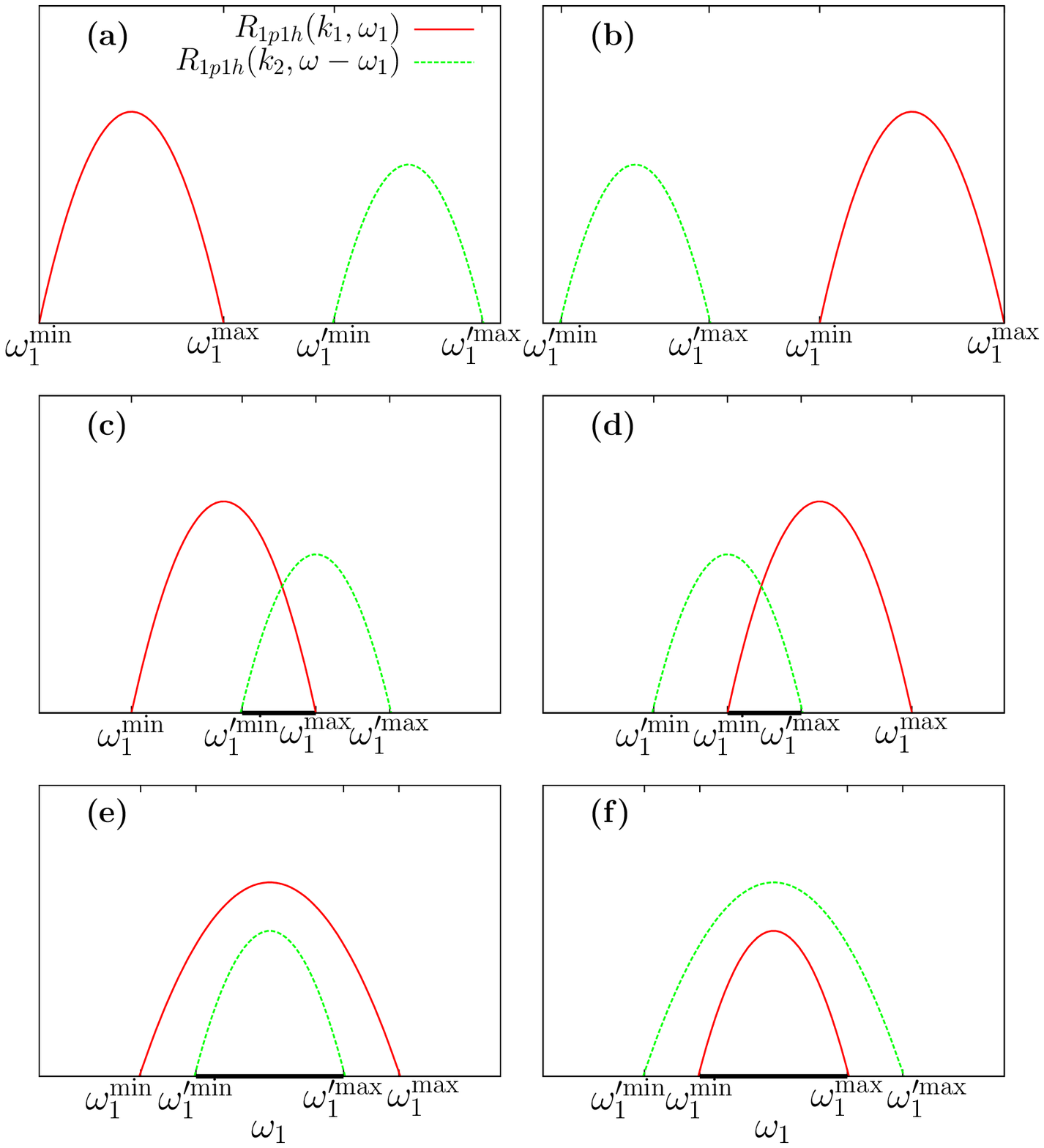}
\caption{Graphs of the two $1p1h$ response functions
  $R_{\rm 1p1h}(k_1,\omega_1)$ and $R_{\rm 1p1h}(k_2,\omega-\omega_1)$
  appearing inside the MCA integral as a function of $\omega_1$. The
  six possible configurations are shown.  In cases a, b they do not
  overlap.  In cases c, d they overlap, and finally, in cases e, f, one
  domain is inside the domain of the other one.  
  }\label{casos}
\end{figure}

\item To obtain the maximum value of $k_1$ we first take into account
that the maximum energy allowed for particle no. 1 is
\begin{equation}
E'_1 \leq \omega + E_F
\end{equation}
{\it i.e.,} all the energy is transferred to particle no. 1, 
initially with $h_1=k_F$.
The corresponding maximum momentum for this on shell particle is
\begin{equation}
p'_1 \leq \sqrt{(\omega+E_F)^2-m_N^2} .
\end{equation}
Therefore the momentum transfer to the first particle is
bound from above by 
\begin{equation}
k_1 = |\np'_1-\nh_1| \leq p'_1 + h_1 
\end{equation}
and 
\begin{equation}
(k_1)_{\rm max} = \sqrt{(\omega+E_F)^2-m_N^2}+k_F .
\end{equation}

\item  Taking into account the fact that the 1p1h elementary response functions
$R_{\rm 1p1h}(k_i,\omega_i)=0 $ for $\omega_i > k_i$, we can restrict the integration over $\omega_1$ between the limits
\begin{equation}
(\omega_1)_{\rm max} = \min (k_1,\omega),
\kern 1cm
(\omega_1)_{\rm min} = \max (0,\omega-k_2) .
\end{equation}

\item For a given value of $k_i$, there is an additional restriction 
for the 1p1h responses, which are zero outside the interval
allowed by the Pauli principle (see  Appendix A for the proof)
\begin{eqnarray}
\omega_i^{\rm min}
& =&
 \sqrt{(k_F-k_i)^2+m_N^2} -E_F  
\kern 1cm (=0  \quad \mbox{if $k_i < 2k_F$})
\nonumber\\
\omega_i^{\rm max}
& =&
\sqrt{(k_F+k_i)^2+m_N^2} -E_F .
\label{omegalimits}
\end{eqnarray}

Therefore the integration limits over $\omega_1$ in Eq. (\ref{MCA}) 
can be further constrained by taking into account that,
inside the integral, two 1p1h response
functions are being multiplied, and both of them 
must be different from zero simultaneously to
contribute. The first response function 
$R_{\rm 1p1h}(k_1,\omega_1)$ is different from zero if 
\begin{equation}
 \omega_1^{\rm min} < \omega_1 < \omega_1^{\rm max} .
\end{equation}
On the other hand, 
$R_{\rm 1p1h}(k_2,\omega-\omega_1)$ is different from zero if 
\begin{equation}
 \omega'_1{}^{\rm min} < \omega_1 < \omega'_1{}^{\rm max} \, ,
\end{equation}
where
\begin{eqnarray}
 \omega'_1{}^{\rm min} = \omega-  \omega_2{}^{\rm max} \\
 \omega'_1{}^{\rm max} = \omega-  \omega_2{}^{\rm min} \, .
\end{eqnarray}
The intersection of the two above intervals defines the final
integration range. This is determined by identifying the six different
possibilities shown in Fig. \ref{casos}.  In the figure we show with
thick lines the resulting integration interval, which depends on the
values of $k_1,k_2$ and $\omega$.  

\end{enumerate}

\section{Electroweak meson-exchange currents}

\begin{figure}[t]
\centering
\includegraphics[width=10cm,bb=110 310 500 690]{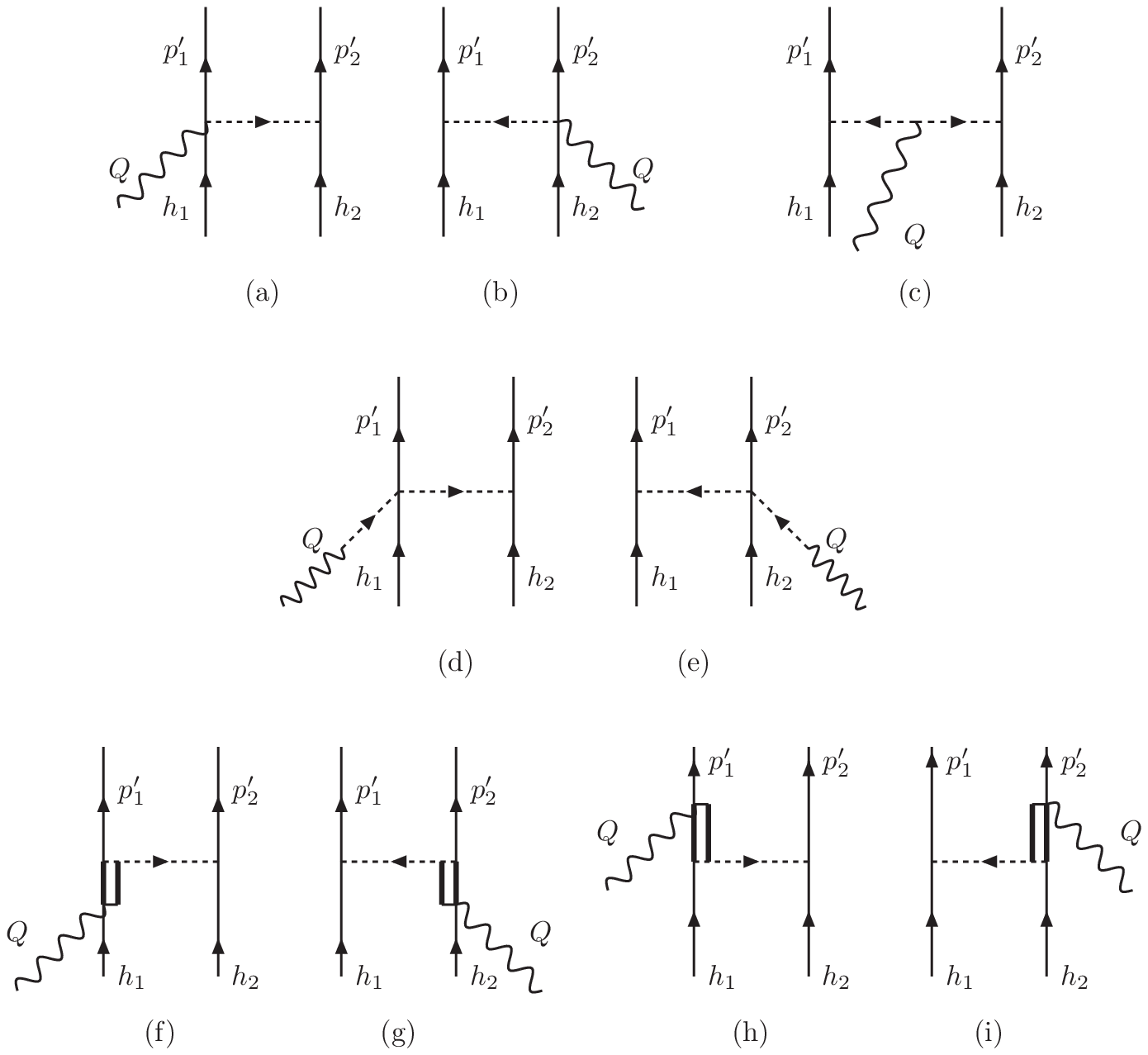}
\caption{Feynman diagrams for the electroweak MEC model used in 
this work.}\label{fig_feynman}
\end{figure}

\begin{figure}[t]
\centering
\includegraphics[width=10cm,bb=180 430 420 690]{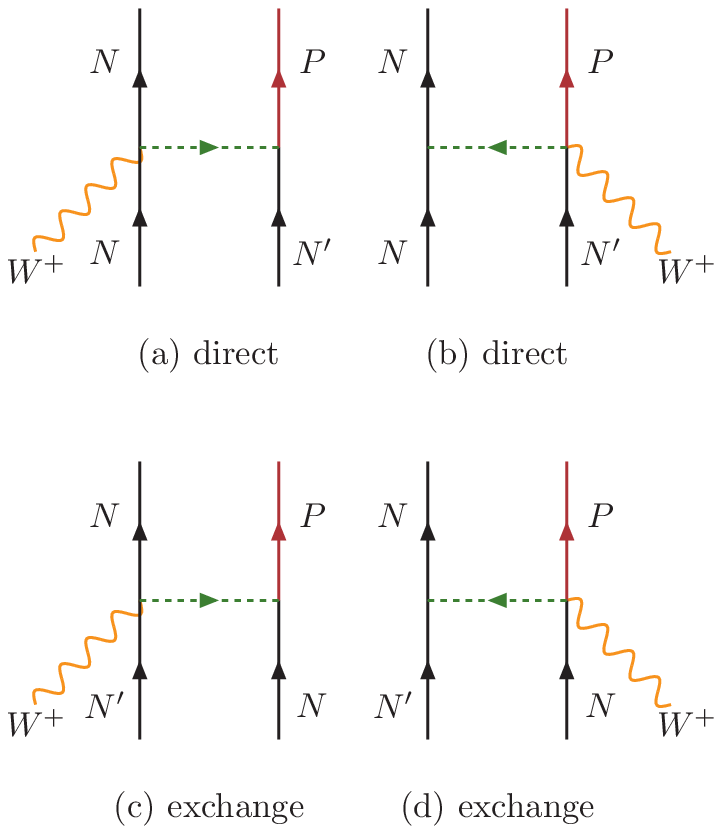}
\caption{Feynman diagrams for 
neutron-proton emission with the seagull current.}\label{seagull1}
\end{figure}

\begin{figure}[t]
\centering
\includegraphics[width=10cm,bb=180 430 420 690]{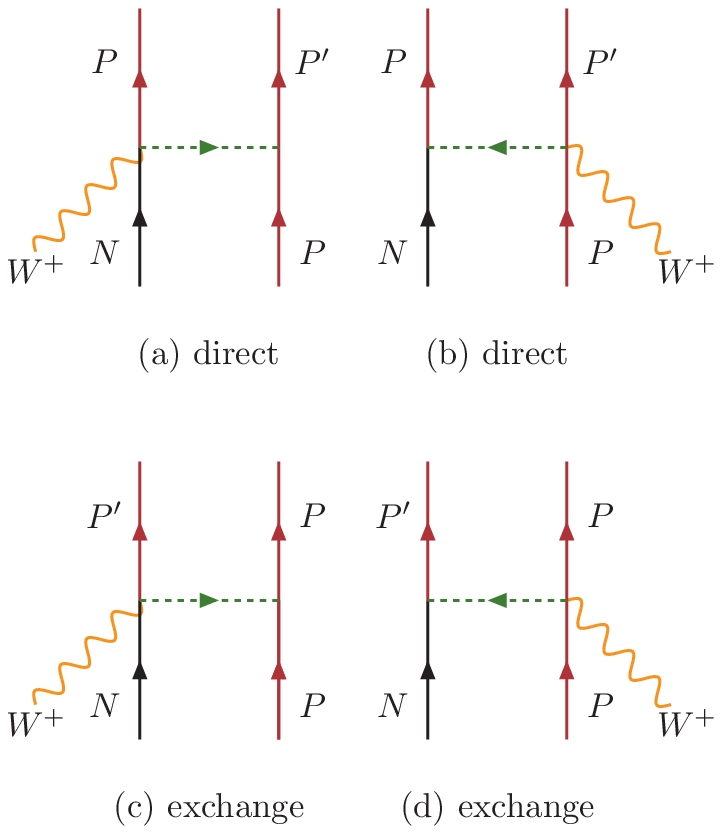}
\caption{Feynman diagrams for 
proton-proton emission with the seagull current.}\label{seagull2}
\end{figure}

In this section we specify a model for the two-body current matrix
elements $j^{\mu}(1',2',1,2)$ entering in the elementary 2p2h
hadronic tensor, Eq. (\ref{elementary}). This will allow us to
investigate the validity of the MCA, by comparing to the full
integration, following the lines of  \cite{Sim14,Rui16}.
The MEC model contains the Feynman diagrams depicted in
Fig.~\ref{fig_feynman}.  
The different  contributions have been taken 
from the pion production model of \cite{Hernandez:2007qq}.
Our MEC is given as the sum of four two-body currents: 
seagull (diagrams a,b), pion in flight (c), pion-pole (d,e) and
$\Delta(1232)$ excitation (f,g,h,i).
Their expressions are given by  
 \begin{eqnarray}
 j^\mu_{\rm sea}&=&
  \left[I_V^{\pm}\right]_{1'2',12}
\frac{f^2}{m^2_\pi}
V_{\pi NN}^{s'_1s_1}(\np'_1,\nh_1) 
 \nonumber\\
 &\times& 
 \bar{u}_{s^\prime_2}(\np^\prime_2)
 \left[ F^V_1(Q^2)\gamma_5 \gamma^\mu
 + \frac{F_\rho\left(k_{2}^2\right)}{g_A}\,\gamma^\mu
   \right] u_{s_2}(\nh_2)
   +
   (1\leftrightarrow2) \,
\label{seacur}
\\
 j^\mu_{\pi}&=& \left[I_V^{\pm}\right]_{1'2',12}
 \frac{f^2}{m^2_\pi}
 F^V_1(Q^2)
V_{\pi NN}^{s'_1s_1}(\np'_1,\nh_1) 
V_{\pi NN}^{s'_2s_2}(\np'_2,\nh_2) 
\left(k^\mu_{1}-k^\mu_{2}\right)
\label{picur}
\\
j^\mu_{\rm pole}
&=&
\left[I_V^{\pm}\right]_{1'2',12}
\frac{f^2}{m^2_\pi}\,
\frac{F_\rho\left(k_{1}^2\right)}{g_A}
\frac{
Q^\mu
\bar{u}_{s^\prime_1}(\np^\prime_1)\Qbar u_{s_1}(\nh_1)
 }{Q^2-m^2_\pi}
V_{\pi NN}^{s'_2s_2}(\np'_2,\nh_2) 
\nonumber\\
&&
+(1\leftrightarrow2)
\label{polecur}
\\
j^\mu_{\Delta}
&=&
\frac{f^* f}{m^2_\pi}\,
V_{\pi NN}^{s'_2s_2}(\np'_2,\nh_2) 
\bar{u}_{s^\prime_1}(\np^\prime_1)
\left\{ 
\left[U_{\rm F}^{\pm}\right]_{1'2',12}
 k^\alpha_{2}
G_{\alpha\beta}(h_1+Q)
\Gamma^{\beta\mu}(h_1,Q)
\right.
\nonumber\\
&& +
\left.
\left[U_{\rm B}^{\pm}\right]_{1'2',12}\; 
k^\beta_{2}
\hat{\Gamma}^{\mu\alpha}(p^\prime_1,Q)
G_{\alpha\beta}(p^\prime_1-Q)
\right\}
u_{s_1}(\nh_1)
+(1\leftrightarrow2)
\label{deltacur}.
\end{eqnarray}
In these equations we have introduced 
the $\pi NN$ vertex function and the pion propagator into the definition 
of the following spin-dependent function:
\begin{equation}
V_{\pi NN}^{s'_1s_1}(\np'_1,\nh_1) \equiv 
\frac{\bar{u}_{s^\prime_1}(\np^\prime_1)\,\gamma_5
 \kbar_{1} \, u_{s_1}(\nh_1)}{k^2_{1}-m^2_\pi} .
\end{equation}
We have also defined the following two-particle isospin operators 
\begin{eqnarray}
I_V^{\pm}  &=& (I_V)_x\pm i (I_V)_y
\\
\Ivec_V  & =&  i \left[\tauvec(1) \times\tauvec(2)\right] ,
\end{eqnarray}
where the $+ (-)$ sign refers to 
 neutrino (antineutrino) scattering.
The forward, 
$U_{\rm F}^{\pm}= U_{Fx}\pm i U_{Fy}$, and
backward, 
$U_{\rm B}^{\pm}= U_{Bx}\pm i U_{By}$, 
isospin transition operators are obtained from the
 Cartesian components defined by
\begin{eqnarray}
U_{\rm Fj}&=&\sqrt{\frac32}
\sum_i\left(T_i T_j^\dagger\right)\otimes
\tau_i
\label{forward}
\\
U_{\rm Bj}&=&
\sqrt{\frac32}
\sum_i\left(T_{j}\, T^\dagger_i\right)\otimes
\tau_i,
\label{backward}
\end{eqnarray}
where $\vec{T}$ is an isovector transition
operator from isospin $\frac32$ to $\frac12$.

In Figs. \ref{seagull1} and \ref{seagull2} we show the
charge-dependent Feynman diagrams contributing to the np and pp
emission channels of CC neutrino scattering in the case of the seagull
current as an example. They involve direct and exchange
contributions. Each current operator contributes analogously. These
figures illustrate the complexity of the elementary 2p2h responses
$r^K$, which here are computed numerically.

We use the $\pi NN$ ($f=1$) and axial ($g_A=1.26$) coupling constants.
The electroweak form factors $F_1^V$ and $F_{\rho}$ in the seagull and
pionic currents are those of the pion production amplitudes of
\cite{Hernandez:2007qq}.  Finally, we use the $\pi N\Delta$ coupling
constant $f^*=2.13$. We also apply strong form factors (not written
explicitly in the MEC) of dipole form in all the $NN\pi$ and $N\Delta
\pi$ vertices.

 For simplicity in this work we only include the dominant terms 
in the weak
$N\rightarrow\Delta$ transition vertex tensor
in the forward current, $\Gamma^{\beta\mu}(P, Q)$
\begin{equation}
\Gamma^{\beta\mu}(P,Q)=
\frac{C^V_3}{m_N}
\left(g^{\beta\mu}\Qbar-Q^\beta\gamma^\mu\right)\gamma_5
+ C^A_5 g^{\beta\mu} .
\end{equation}
We have kept only the $C_3^V$ and $C_5^A$ form factors and neglected the 
smaller contributions of the others.
They are taken from \cite{Hernandez:2007qq}.
On the other hand, for the backward current, the vertex tensor is
\begin{equation}
\hat{\Gamma}^{\mu\alpha}(P^\prime, Q)=\gamma^0
\left[\Gamma^{\alpha\mu}(P^\prime,-Q)\right]^{\dagger}
\gamma^0 \, .
\end{equation}

Finally
the $\Delta$-propagator
takes
into account the finite decay width of the $\Delta\,(1232)$
by the prescription
\begin{equation}\label{delta_prop}
 G_{\alpha\beta}(P)= \frac{{\cal P}_{\alpha\beta}(P)}{P^2-
 M^2_\Delta+i M_\Delta \Gamma_\Delta+
 \frac{\Gamma^2_{\Delta}}{4}} \, .
\end{equation}
In this work we consider both the real and imaginary parts of the denominator 
of this propagator.
The projector  ${\cal P}_{\alpha\beta}(P)$ over 
spin-$\frac32$ on-shell particles is given by
\begin{eqnarray}
{\cal P}_{\alpha\beta}(P)&=&-(\Pbar+M_\Delta)
\left[g_{\alpha\beta}-\frac13\gamma_\alpha\gamma_\beta-
\frac23\frac{P_\alpha P_\beta}{M^2_\Delta}\right.
\nonumber\\
&+&\left.
\frac13\frac{P_\alpha\gamma_\beta-
P_\beta\gamma_\alpha}{M_\Delta}\right].
\end{eqnarray}
We do not take into account possible off-shellness effects in this projector.

\section{Treatment of the $\Delta$-propagator}

Using an average value for the hole momenta  
inside the MCA integral will be valid only if
 the elementary 2p2h response functions 
depend slowly on $\nh_1$ and $\nh_2$.
This is not the case for the forward $\Delta$ diagram, 
which presents a sharp
maximum  due to the pole structure 
of the $\Delta$ propagator,
\begin{equation}
G_{\Delta}(H+Q)
\equiv
\frac{1}{(H+Q)^2-
 M^2_\Delta+i M_\Delta \Gamma_\Delta+
 \frac{\Gamma^2_{\Delta}}{4}} \, ,
\label{denominator}
\end{equation}
where $H^\mu=(E_{\nh},\nh)$ is the four-momentum of the hole.  Taking
an average value for the momentum instead of computing the full
integral modifies the pole position, and this distorts the shape and
strength of the 2p2h $\Delta$ peak. Due to this effect, the present
MCA approach is not as accurate as in the low-energy region far from
the $\Delta$ peak. In this work we compare the results of different 
methods to improve the description of the $\Delta$ peak:

\begin{enumerate}
\item the full $\Delta$ propagator, Eq. (\ref{denominator}),  but using the
corresponding average momentum; 

\item the Fermi-averaged or frozen $\Delta$ propagator;

\item the cone-averaged $\Delta$ propagator.
\end{enumerate}

These averaged propagators are explained below.

\subsection{The frozen $\Delta$  propagator}

This is a propagator averaged over the momentum distribution of
  the Fermi gas. This produces a smearing of the $\Delta$ peak. This was
  proven to be an excellent approximation in the case of the frozen
  approximation of \cite{Rui17}, where the momentum of the hole
  was approximated by zero. Therefore this choice amounts to compute the average integral, by
taking the non-relativistic
limit for the energies of the hole ($E_{\nh}\simeq m_N$), 
\begin{eqnarray}
G_{\rm frozen}(Q)
&=&
\frac{1}{\frac43 \pi k^3_F}
\int 
 \frac{d^3h \;\theta(k_F-\left|\nh\right|)}{a-2\,\nh\cdot\nq+ib} \, ,
\\
&=& 
\frac{1}{\frac43 \pi k^3_F} 
\frac{\pi}{q}\left\{ \frac{\left(a+ib\right)k_F}{2q}
\right.
\label{averaged_denom}\\
&&
\kern -1cm \mbox{}+\left.
\frac{4q^2k^2_F - (a+ib)^2}{8q^2}
\ln\left[\frac{a+2k_Fq+ib}{a-2k_Fq+ib}\right]
\right\} \, ,
\nonumber
\end{eqnarray}
where the parameters $a,b$ are defined by  
\begin{eqnarray}
\kern -0.8cm 
a &\equiv& m^2_N+Q^2+2m_N(\omega+\Sigma)-M^2_\Delta+
\frac{\Gamma^2}{4}
\label{afrozen}\\
\kern -0.8cm 
b &\equiv& M_\Delta \Gamma
\label{bfrozen} \,.
\end{eqnarray}
They depend on two parameters, $\Gamma$ and $\Sigma$, that correspond
to an effective width and shift of the smeared $\Delta$ peak.  These
parameters are adjusted to reproduce the full results with the 7D
integral, and depend on the momentum transfer $q$.  They are given in
Table \ref{table1}.  Note that $\Gamma$ is slightly different from the
values fitted in \cite{Rui17}. 
The latter is because in this fit we use the MCA instead 
of the frozen approximation
used in \cite{Rui17}.

\begin{table}[h]
\centering
\begin{tabular}{|c|c|c|c|}\hline 
$q$ (MeV/c) & $\Sigma$ (MeV) & $\Gamma$ (MeV) & $\Gamma_c$ (MeV) \\\hline 
300 & 20 & 110 & 145 \\\hline
400 & 65 & 135 & 138 \\\hline
500 & 65 & 125 & 134 \\\hline 
800 & 80 & 100 & 128 \\\hline 
1000 & 100 & 80 & 125 \\\hline 
1200 & 115 & 60 & 123 \\\hline 
1500 & 150 & 20 & 120 \\\hline 
2000 & 150 & 0  & 105  \\\hline
\end{tabular}
\caption{Values of the free parameters of the Fermi-averaged
$\Delta$-propagator for
different values of the momentum transfer $q$.}\label{table1}
\end{table}

\begin{figure}[t]
\begin{center}
\includegraphics[width=9cm, bb=200 540 300 630]{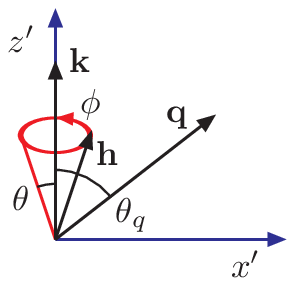}
\caption{ Scheme defining the geometry 
for the cone-averaged integration of the $\Delta$ propagator. The 
$z'$ axis is defined by  the direction of the partial momentum transfer
$\nk$ to the hole $\nh$. The modulus $h$ and cone angle $\theta$ 
are fixed by the kinematics of the 1p1h response. 
The azimuthal angle 
$\phi$ is undetermined and is integrated around the $z'$ axis 
for the  average.
The result depends on the cone geometry and 
on the angle $\theta_q$ between 
$\nk$ and the momentum transfer $\nq$ as well. }
\label{anguloscono}
\end{center}
\end{figure}

\subsection{Cone-averaged $\Delta$-propagator}

In the present MCA approach the moduli of the momenta $h_i$, and the
angles $\theta_i$ between $\nh_i$ and $\nk_i$ are fixed, and therefore
the vectors $\nh_i$ belong to the cones shown in
Fig. \ref{conos}. However the azimuthal angles in the cones are not
determined, and here we choose different prescriptions to fix them.
By doing that, the $\Delta$-peak position is altered with respect to the exact
value obtained in the full 7D integral. The cone-averaged propagator
introduced here is defined by smearing the $\Delta$ propagator by
averaging only in the cone 
instead of averaging over the full Fermi
sea:
\begin{equation}
G_\Delta(H+Q) \rightarrow 
G_{\rm cone}
\equiv
\frac{1}{2\pi}
\int 
 \frac{d\phi}{a_c-2\,\nh\cdot\nq+ib_c} \, ,
\end{equation}
where the integration variable $\phi$ 
is  the azimuthal angle around the cone shown in Fig. \ref{anguloscono},
$\nk$ is the partial 
momentum transfer to the hole $\nh$ and $\nq$ is the total momentum transfer.
The cone-averaged propagator depends on the 
parameters in the denominator, $a_c,b_c$,  defined by  
\begin{eqnarray}
\kern -0.8cm 
a_c &\equiv& m^2_N+Q^2+2E_h\omega-M^2_\Delta+
\frac{\Gamma_c^2}{4}
\label{acono}\\
\kern -0.8cm 
b_c &\equiv& M_\Delta \Gamma_c
\label{bcono} \,.
\end{eqnarray}
These cone parameters are independent 
of $\phi$. The $\phi$ dependence
is hidden in the scalar product $\nh\cdot\nq$. To obtain this dependence we
use Fig. \ref{anguloscono} in the rotated coordinate system, $x'z'$, in the
scattering plane, where the $z'$ axis points along $\nk$.
In this system the scalar product is
\begin{equation}
\nh\cdot\nq=hq(\sin\theta\sin\theta_q\cos\phi+\cos\theta\cos\theta_q) \, ,
\end{equation}
where $\theta$ is the angle between $\nh$ and $\nk$, defining the cone,
and $\theta_q$ is the angle between $\nk$ and $\nq$.
The cone-averaged propagator can be expressed as the integral 
\begin{equation}
G_{\rm cone}
=
\frac{1}{2\pi}
\int 
 \frac{d\phi}{W-C\cos\phi} \, ,
\end{equation}
with
\begin{eqnarray}
W & = & a_c+ib_c - hq \cos\theta\cos\theta_q
\\
C & = & hq \sin\theta\sin\theta_q .
\end{eqnarray}
Note that $C>0$ and $W$ is a complex number. The above integral is
performed analytically in Appendix B.

The values of the effective width in the cone-averaged
propagator, $\Gamma_c$, are tabulated as a function of $q$ in Table
\ref{table1}.

\begin{figure}[tp]
\begin{center}
\includegraphics[width=8cm, bb=170 310 450 780]{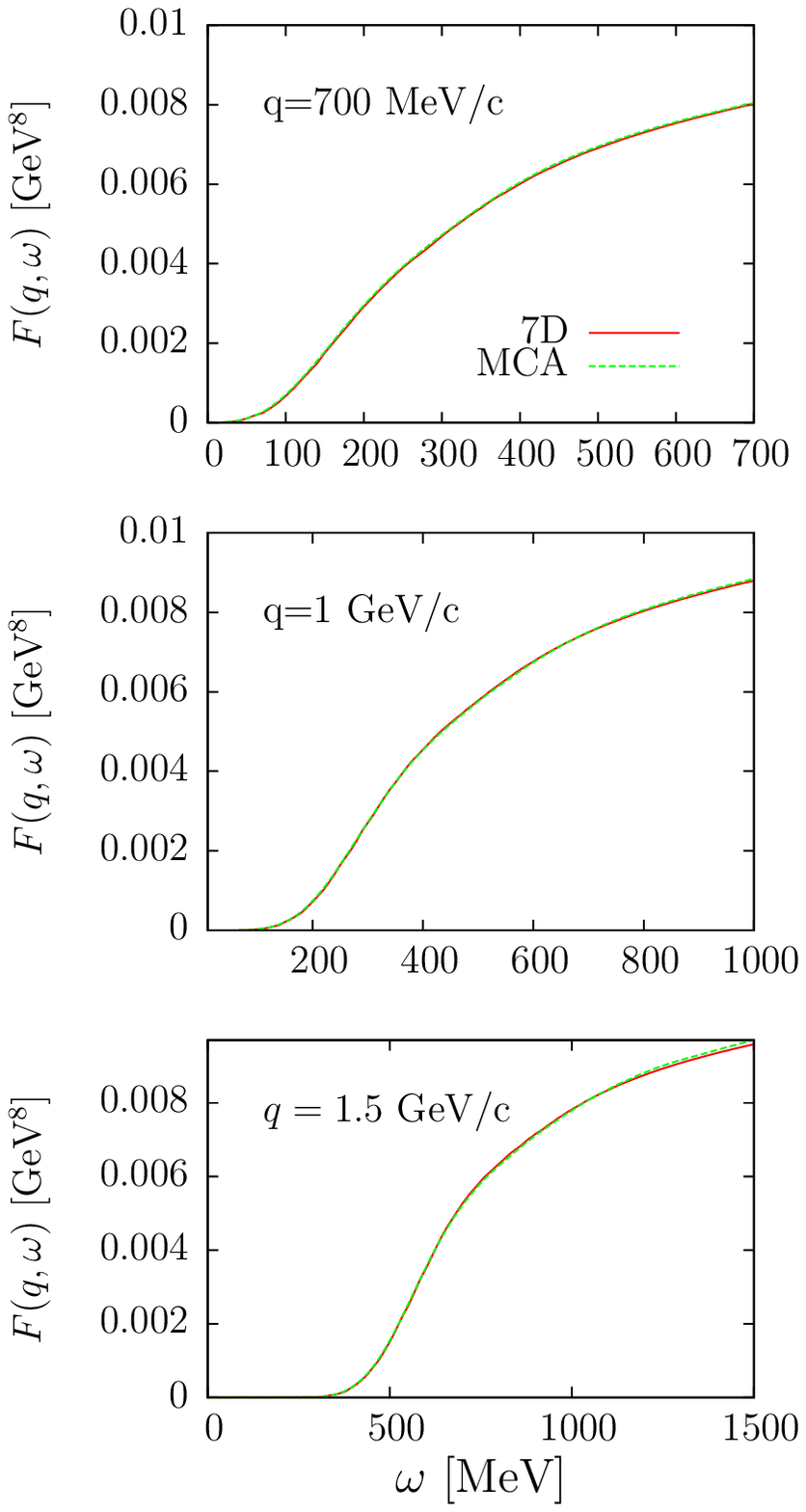}
\caption{ Phase space function computed for three values of the
  momentum transfer, as a function of $\omega$. 
  Results obtained with the full 7D integral are 
  compared to the MCA approach with only 3D integrations. 
 }
\label{fasico1}
\end{center}
\end{figure}

\section{Results}\label{sec_results}

In this section we present results for the 2p2h response functions
for inclusive neutrino scattering. In this work we do not provide
 comparisons with the experimental data. This requires one to
describe simultaneously the quasielastic and inelastic (including pion
emission) channels. In previous works \cite{Meg16a,Meg16b} we have
provided this comparison within the superscaling approach plus a MEC
model derived from the one used in the present work. This model
describes the $(e,e')$ cross section of $^{12}$C and the global set of
neutrino scattering quasielastic without pions (CC0$\pi$) measurements
made in the neutrino accelerator experiments. Therefore the model we
are starting with is realistic for describing two-nucleon emission
with neutrinos for the kinematics of interest.

In particular we investigate the validity of the MCA approach
presented so far, by evaluating the 2p2h response functions and
comparing with the full results obtained with the 7D integration.  The
interest of this investigation is to determine the consistency between
different approaches to the 2p2h emission channel, namely the model of
\cite{Nie11,Gra13}.  This is a first step towards the reconciliation
between apparently different approaches. This study is a necessary
step forward in reducing the systematic errors in the oscillation
parameters coming from the theoretical uncertainties.  Moreover it is
interesting by itself to find alternative approximations that allow a
reduction of the computational time of the two-nucleon emission
without large loss of numerical precision.  Finally we remark that
the MCA
allows us to make an easier connection between our results and the
predictions provided by other authors \cite{Nie11}.

 We consider the case of the nucleus ${}^{12}$C, and, unless otherwise
 stated, the Fermi momentum is chosen to be $k_F=228$ MeV/c.  We show
 results for several kinematics in the range of momentum transfer
 between 300 MeV/c and 2 GeV/c, of interest for the neutrino
 oscillation experiments.

\begin{figure}[tp]
\begin{center}
\includegraphics[width=11cm, bb=50 220 550 780]{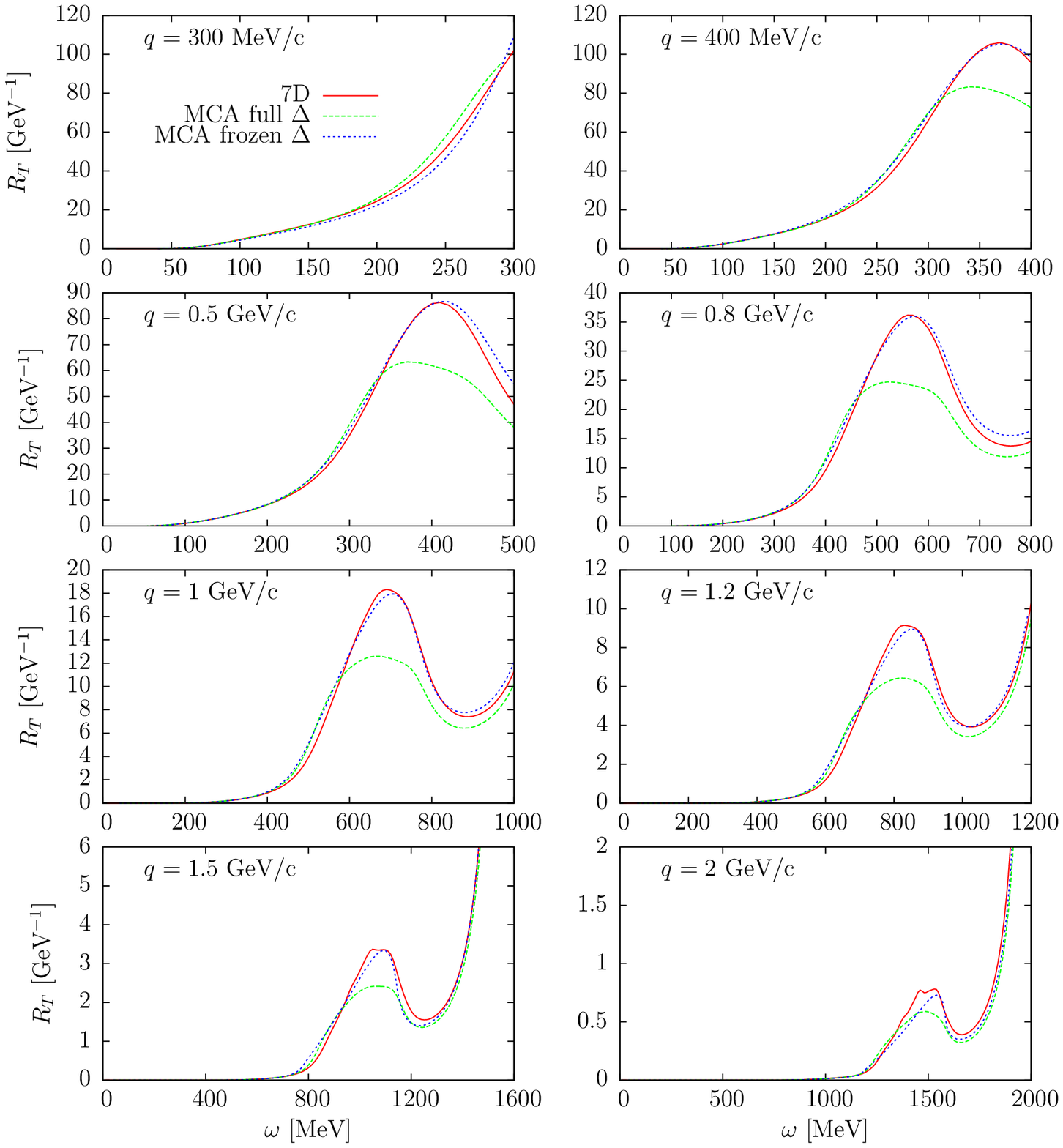}
\caption{ Transverse 2p2h response function as a function of $\omega$
  for several values of the momentum transfer. The results of the RFG
  with the full 7D integration are compared to the MCA with the full
  $\Delta$ propagator and with the Fermi-averaged or frozen $\Delta$
  propagator.  }
\label{fig3}
\end{center}
\end{figure}

We start by computing 
 the phase-space function obtained from Eq. (\ref{hadronic}) for
$r^K=1$, except for a constant factor. This is a universal function for the RFG that only depends 
on the kinematics of the 2p2h excitations and not on the 
MEC model:
 \begin{eqnarray}
F(q,\omega)
&=&
\int
d^3p'_1
d^3h_1
d^3h_2
\frac{m_N^4}{E_1E_2E'_1E'_2}
\Theta(p'_1,h_1)\Theta(p'_2,h_2)
\nonumber \\ 
&&
\delta(E'_1+E'_2-E_1-E_2-\omega) .
\label{espaciofasico}
\end{eqnarray}
This universal function is well established and was fully studied in
 \cite{Sim14,Sim14b}. Therefore the comparison with this function
is a required consistency check in any calculation of the 2p2h
response, and furthermore it provides a valuable precision check of
the multidimensional integrations. In fact the phase space determines
the global behavior of the 2p2h responses, on top of the additional
modifications introduced by the particular model of two-body current
operator. The main one is produced by the $Q^2$ dependence of the
electromagnetic form factors and the structure of the different
diagrams, which are dominated by the forward $\Delta$ excitations. The remaining
$(q,\omega)$ dependence of the MEC diagrams is found to be smoother.

The results in Fig. \ref{fasico1} confirm numerically that the MCA
approach treats exactly the phase space in all energy regions,
obtaining essentially the same results except for numerical errors in
the integration procedure. This comparison also allows us to determine
the optimal integration steps in the MCA.

In Fig. \ref{fig3} we compare the transverse 2p2h response function
of the RFG computed 
by performing the full 7D integration with the MCA results
for two different prescriptions for the $\Delta$ propagator.  Using the
full $\Delta$ propagator in the MCA produces a peak that is about
$30\%$ smaller than the full result, and slightly shifted and
distorted. This is due to the approximation made for the averaged
momentum of the hole, and to the chosen values of the azimuthal angles
in the cones shown in Fig.  \ref{conos}. These results have been
obtained for $\alpha_1=\alpha_2=0$, corresponding to both holes
contained in the scattering plane on the same side of the cone, and
corresponding to choosing the $+$ sign in Eq. (\ref{ui}). Therefore the
position and the value of the maximum due to the forward $\Delta$
propagator is altered by the average in the MCA.  This problem is
dealt with in this example by using the smeared frozen $\Delta$ propagator
averaged over the RFG momentum distribution, as also shown in
Fig. \ref{fig3} with dotted lines, which are quite similar to the full
results, after fitting the two parameters of the effective width and
shift, $(\Gamma,\Sigma)$, shown in Table \ref{table1}.  Note that for low
transferred energy,  far from the resonance position, and especially at
threshold, all the results coincide independently of the $\Delta$-propagator
treatment. Thus, globally, the MCA results using the frozen
$\Delta$ propagator are quite satisfactory.

Nonetheless the quality of the agreement relies on using the fitted
values for the width and shift, which 
depend strongly on $q$. However, notice that this procedure may limit the predictability of the present approach in so far as its reliability is
linked to the knowledge of the full results.

\begin{figure}[tp]
\begin{center}
\includegraphics[width=7cm, bb=150 350 450 780]{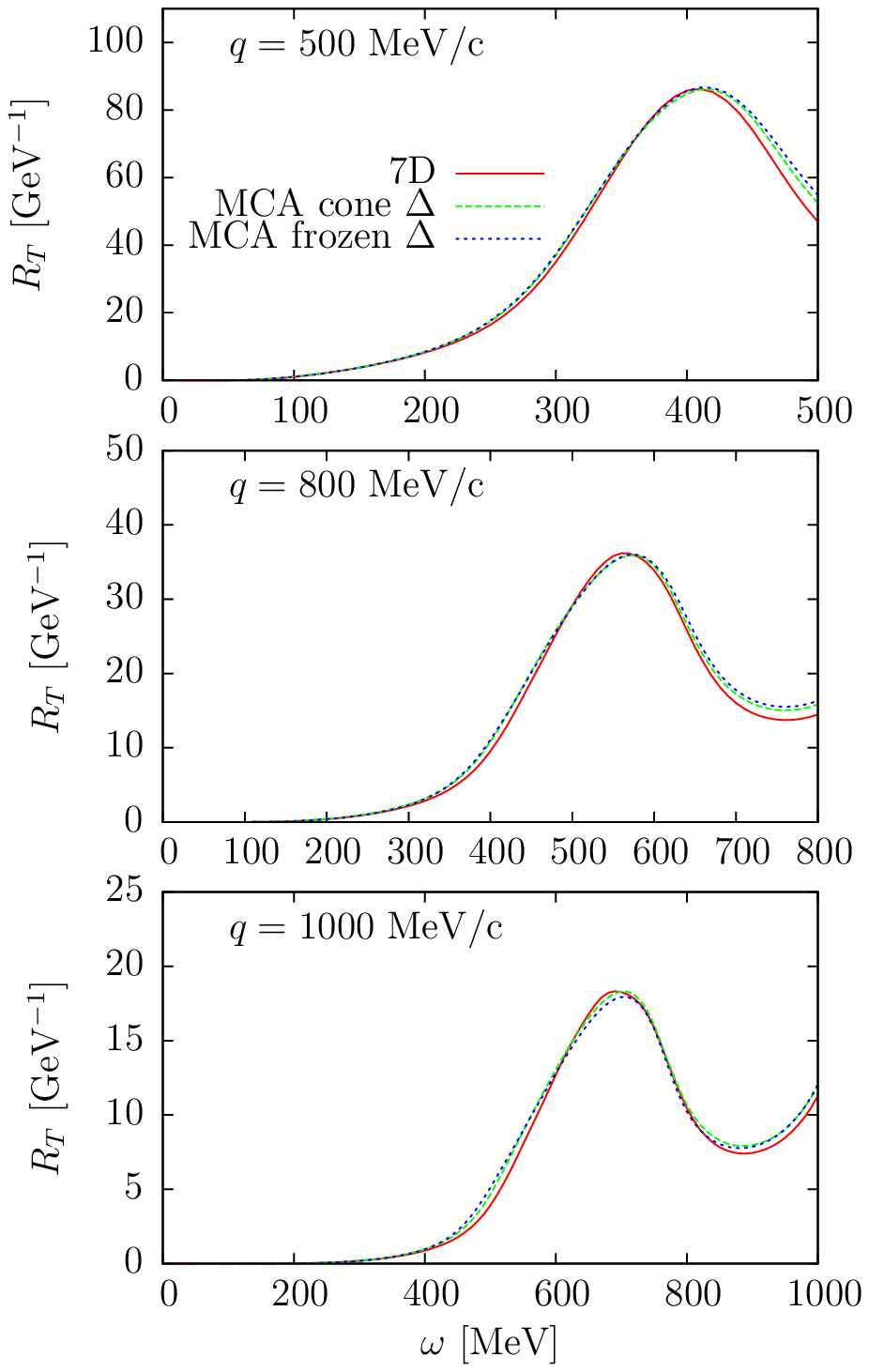}
\caption{ Transverse 2p2h response function as a function of $\omega$
  for several values of the momentum transfer. The results of the RFG
  with the full 7D integration are compared to the MCA with the cone averaged
  $\Delta$ propagator and with the frozen $\Delta$
  propagator.  }
\label{fig4}
\end{center}
\end{figure}

An alternative solution is to use the cone-averaged propagator, which only
depends on one parameter $\Gamma_c$ with a milder dependence on $q$, as
shown in Table \ref{table1}, which oscillates between 145 and 105 MeV
in the $q$ range considered, more or less around the free $\Delta$
width, $\Gamma_c = 125 \pm 20$.  On the other hand, the frozen width
$\Gamma$ changes in the larger range between 135 and zero.  Results
with the cone-averaged $\Delta$ propagator are shown in
Fig. \ref{fig4}. They are similar to the exact results and the quality
of this approximation is at least as good as using the frozen $\Delta$
propagator. However the cone-averaged approximation has the advantage
of having only one free parameter, the $\Gamma_c$ width, which
is also closer to the $\Delta$ free width.
The agreement with the full results is remarkable
 in  the full range of momentum transfer explored in this work.

From now on all the results of the MCA shown will be
calculated using the cone-averaged $\Delta$ propagator, 
with the parameters displayed in the last column of Table 
\ref{table1}.

\begin{figure}[tp]
\begin{center}
\includegraphics[width=11cm, bb=50 480 550 780]{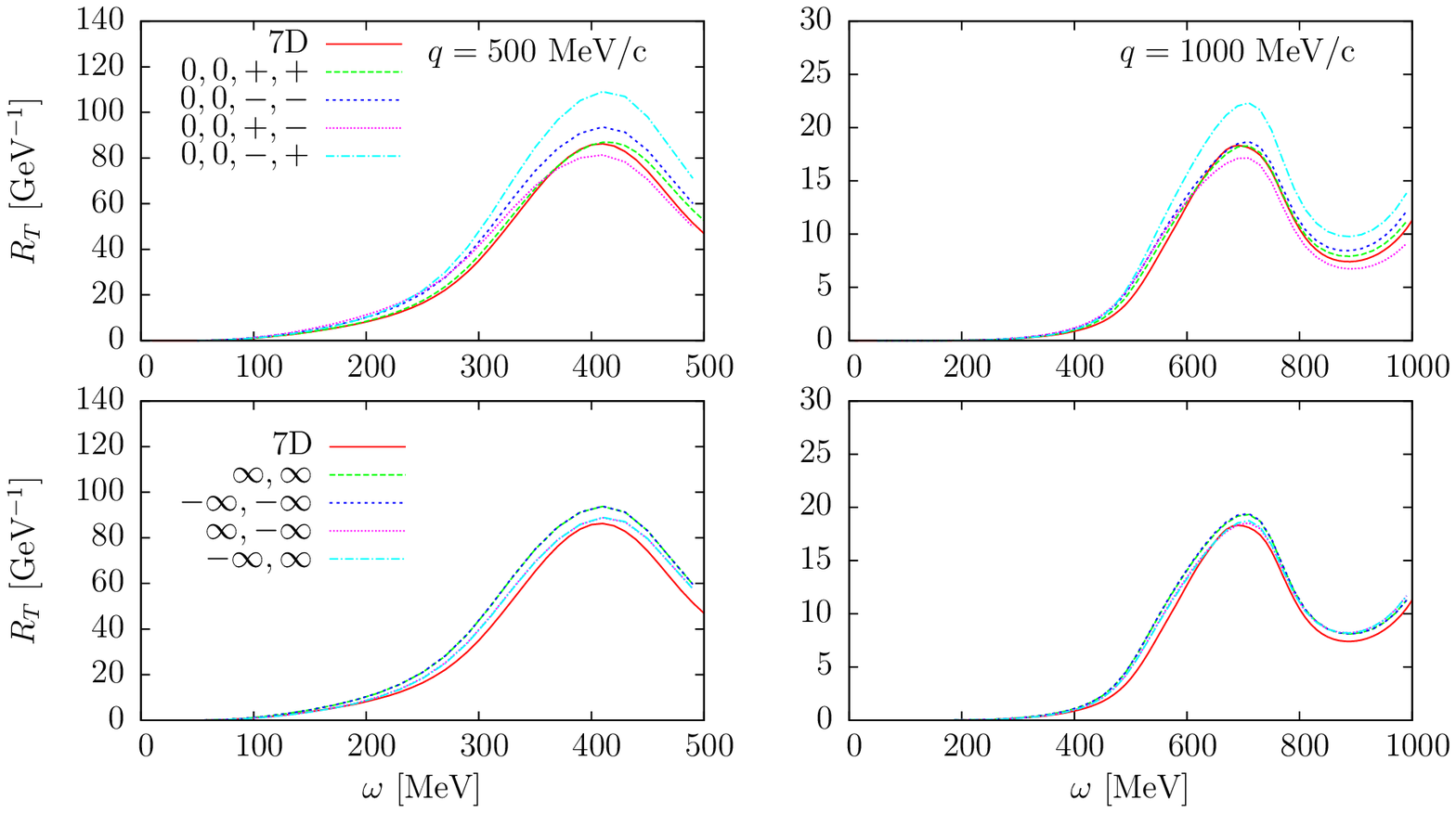}
\caption{2p2h transverse response function plotted against
  transferred energy, $\omega$, for different $q$-values ($q=500$
  MeV/c on the left side and $q=1000$ MeV/c on the right one) and for
  different average momenta orientations, labelled by the keys
  defined in Fig.
  \ref{configurations}.}
\label{fig5}
\end{center}
\end{figure}

\begin{figure}[h]
\subfloat[Configurations for initial holes' momenta
on the scattering plane, labelled by the key indicating
their ($\alpha_1,\alpha_2$) values and their signs
corresponding to Eq.~(\ref{ui}).\label{onplane}]{%
\includegraphics[width=5.5cm, bb=200 400 400 660]{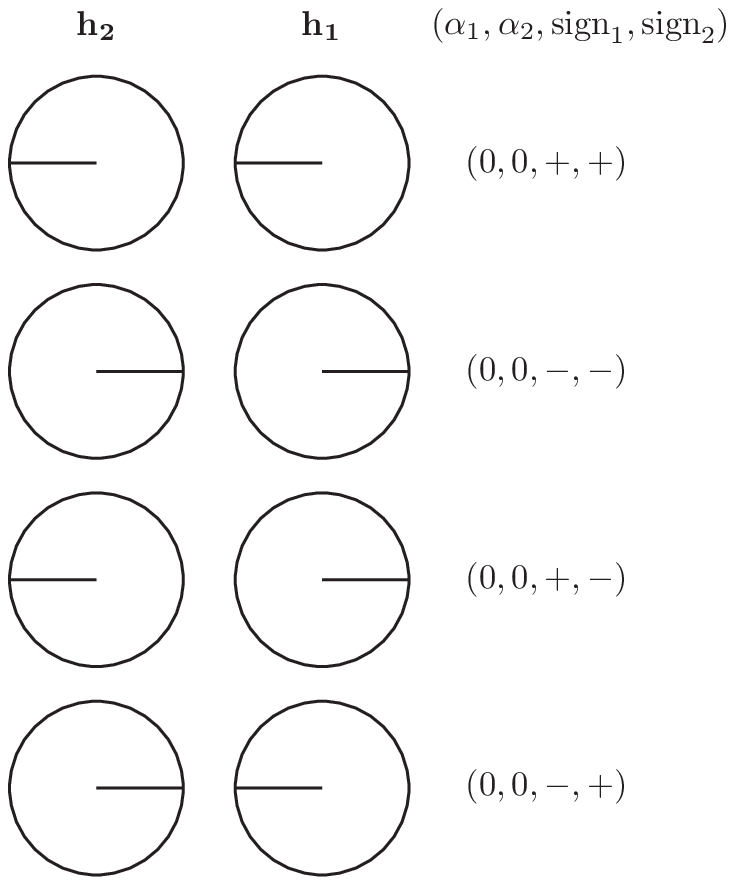}
}
\hfill
\subfloat[Configurations for initial holes' momenta
out of the scattering plane, labelled by the key indicating
their ($\alpha_1,\alpha_2$) values and their signs
corresponding to Eq.~(\ref{ui}).\label{outofplane}]{%
\includegraphics[width=5.5cm, bb=200 400 400 660]{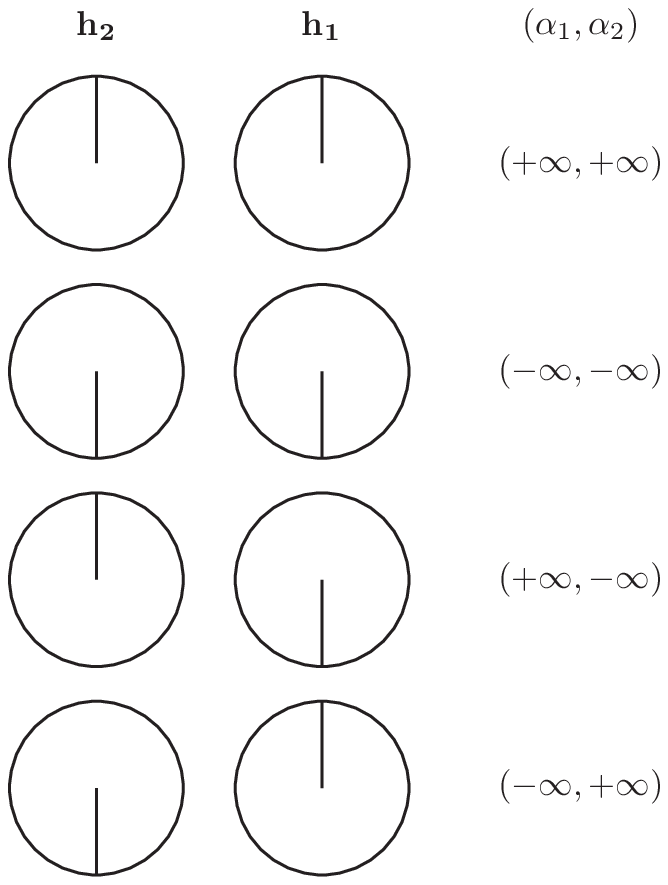}
}
\caption{Different initial holes' configurations considered in
this work.}
\label{configurations}
\end{figure}

In Fig. \ref{fig5} we study the dependence of the MCA results on the
chosen value for the azimuthal angles of the initial hole momenta. As
was shown in Fig. \ref{conos}, each averaged momentum $\nh_i$ is
arranged in the lateral surface of the corresponding cone depicted in
the figure. The precise value of the azimuthal angle measured with
respect to $\nk_i$ (the cone axis) is determined by the parameters
$\alpha_i$ in Eq.~(\ref{ui}) and the sign of the unit vector
$\unit_i$. Any combination of pairs of azimuthal angles between 0 and
$2\pi$ is possible. In Fig. \ref{fig5} we show different pairs of
choices and compare the corresponding $T$ responses.  In
Fig. \ref{configurations} we show the configurations chosen for this
study, viewed from the cone bases.  In the (a) case the two holes are
in the scattering plane but we change their positions among the two
possible sides of the cones, corresponding to $\alpha_i=0$, and sign
$\pm$ for $\unit_i$. In the (b) case the hole momenta are out of the
scattering plane with the maximum angle allowed, corresponding to
$\alpha_i= \pm \infty$.  As we see, the results depend mildly on the
different choices, and we can identify configurations which are
particularly stable with little dependence on the chosen
values.  In particular, the MCA results computed for
$\alpha_i=\pm\infty$ are all 
quite similar to the full 7D
response function, all of them being in an interval around $\sim
5\%$ above the full result.  The same can be said for the $(0,0,-,-)$
configuration, while the worst results occur for $(0,0,-,+)$.  This is
related to the fact that we are constraining the two $\nh_i$ momenta to
be the closest to the momentum transfer $\nq$, as can be seen in
Fig. \ref{configurations}. This restricts the possible arrangements of
the hole pairs in the average momenta approximation. We conclude that
the maximum uncertainty of the MCA methods comes from this choice.
Note that 
of all these configurations, only the case $(0,0,+,+)$ was
fitted to the full response, while the others are not fitted, and are
computed with $\Gamma_c$ parameter fixed to this case.

\begin{figure}[tp]
\begin{center}
\includegraphics[width=9cm, bb=150 360 450 800]{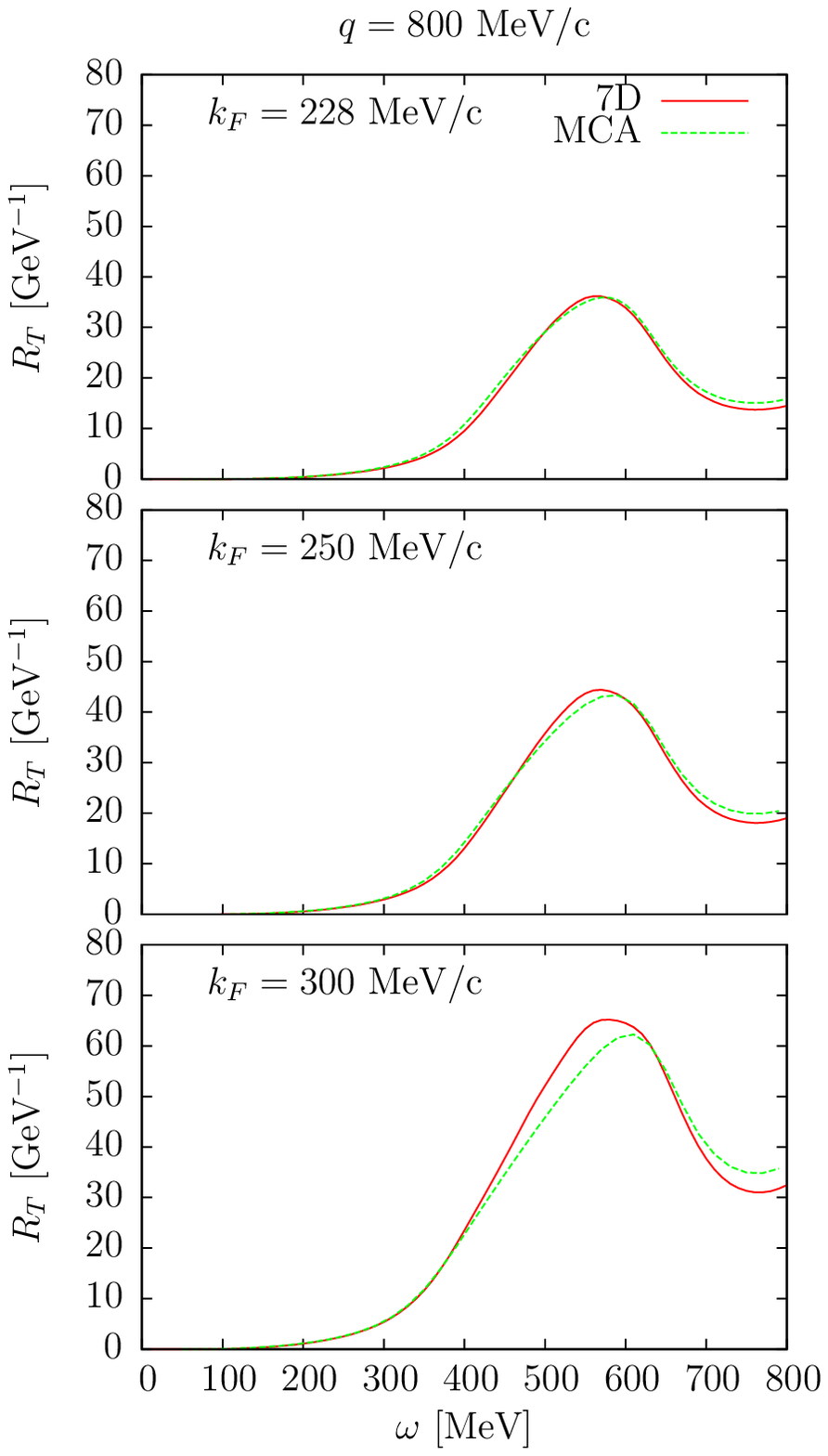}
\caption{2p2h transverse response function of $^{12}$C, 
for $q=800$ MeV/c, 
plotted against
   $\omega$, for different values of the Fermi momentum $k_F$.
}
\label{fig6}
\end{center}
\end{figure}

In Fig. \ref{fig6} we show an example of how the present MCA results
behave when the value of the Fermi momentum is changed. Note that in
all cases the nucleus considered is $^{12}$C, but the Fermi momentum
is increased up to $k_F=300$ MeV. In a different nucleus, these results
should be rescaled with the number of particles in addition to the
increase observed in the figure when $k_F$ is enlarged. In fact the
$R_T$ 2p2h response per nucleon is almost a factor of two
when $k_F$ changes from 228 to 300 MeV/c.
 More precisely, the ratio between the maxima of the two responses is 
about 1.75, in agreement with the results of \cite{Amaro:2017eah}, where it was shown that the 2p2h response functions scale as $k_F^2$.
As we see the MCA results
are quite stable in this range of $k_F$, yet the parameter $\Gamma_c$
was fitted for one particular $k_F=228$ MeV/c. And this is why the
results worsen a bit for higher $k_F$. These results show that the
MCA approach can be applied to heavier nuclei. Although not presented here, 
the present formalism can be extended with small
changes to the case of $N\ne Z$ nuclei, with different Fermi momenta for protons
and neutrons. This will allow one to compute the 2p2h cross sections for
neutrino scattering from detectors made of different nuclei (typically C, O, Ar) such as the ones used in ongoing neutrino experiments.

\begin{figure}[tp]
\begin{center}
\includegraphics[width=11cm, bb=50 360 550 800]{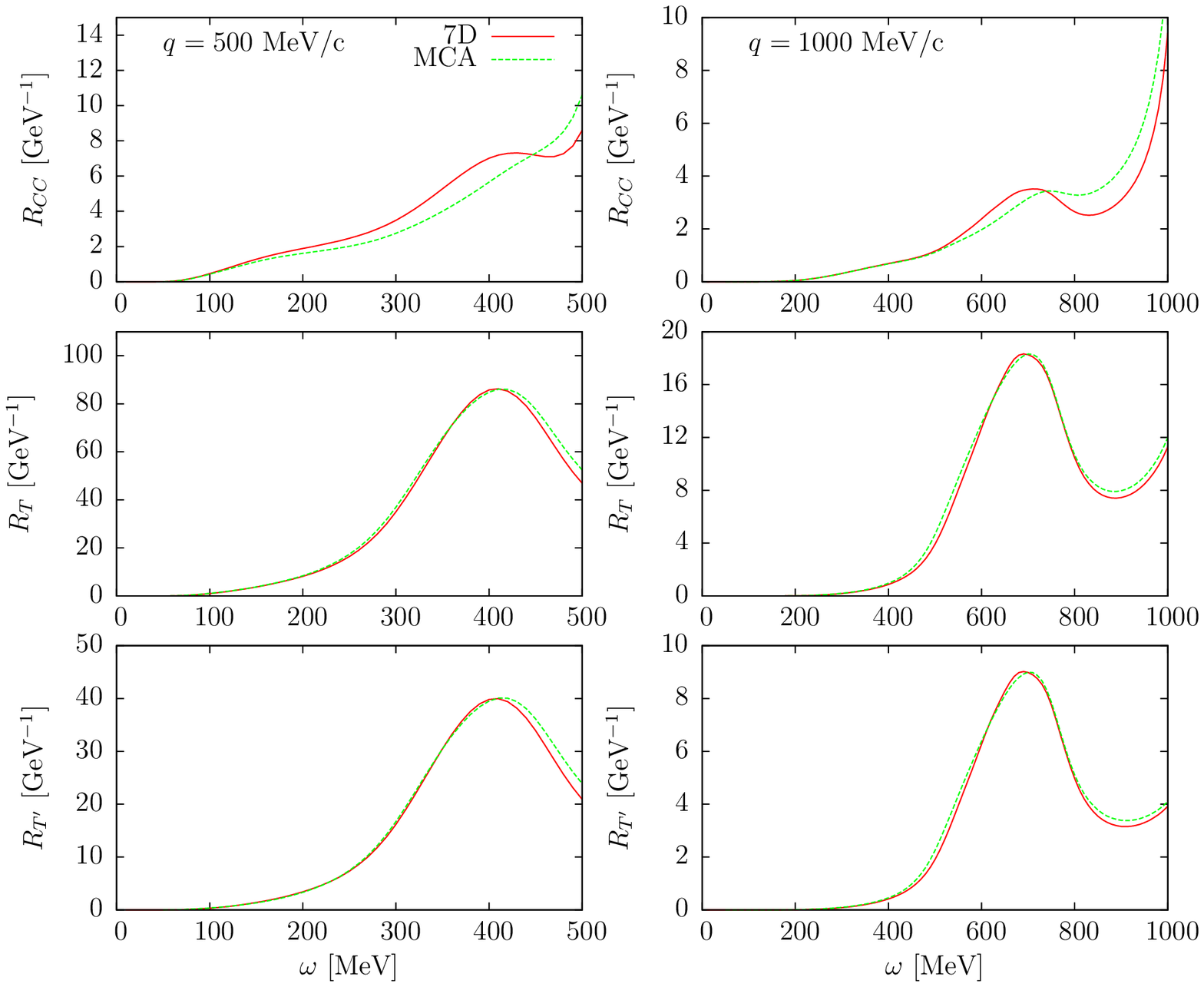}
\caption{Comparison of the MCA and full results for the 
 $CC$, $T$ and $T'$ 2p2h response functions of $^{12}$C, 
and for two values of the momentum transfer.
}
\label{fig7}
\end{center}
\end{figure}

Finally, in Fig. \ref{fig7} we show that the MCA approach works quite well
for the 2p2h response functions of the different kinds, although the
parameters have been fitted only for the $T$ response. The worst
agreement occurs for the small $R_{CC}$, which partially cancels with
the $CL$ and $LL$ responses. The main contribution to neutrino
scattering thus comes from the $T$ and $T'$ responses, which are well
described in the MCA approach.

\section{Conclusions}\label{sec_conclusions}

In this work we have introduced a simplified approach, the modified
convolution approximation (MCA), which allows one to write the 2p2h response
functions as a convolution of two single-particle response functions
weighted with the average 2p2h elementary responses.  The approach
treats exactly the kinematics of the 2p2h excitation
and therefore it works the best for low energy transfers.
The resulting
approximation allows one to reduce the number of integrals from seven to
three dimensions, with a considerable saving in computational effort.
Our formalism includes the interference between direct and exchange diagrams.

After introducing the general formalism in this approach, we have
tested its quality and precision by comparing with the full
results using a specific model of relativistic two-body MEC operators.
The approach works well when an appropriate smearing of the $\Delta$
propagator, obtained by averaging it over the hole momenta, is used.  This approximation
requires to choose the specific direction of the average momenta for
the initial nucleons in the current matrix elements, which is
contained over the surface of a cone.  We have found that a simple
averaged propagator over the azimuthal angle of the hole momentum
around the cone gives quite good results for all the values of the
kinematics.  The ambiguities in the model related to the prescription
for the direction of the hole over the cone surface are found to be
mild. 

The MCA presented here can also be considered as the natural
generalization of the pioneering non-relativistic formalism of
 \cite{Van81} to the relativistic case, which requires to add an
additional integral over the energy $\omega_1$ transferred to the first nucleon.

Furthermore the present approach provides an integral representation of
the 2p2h responses which is similar to the model of Valencia
\cite{Nie11} and therefore this can yield a comparison of the
compatibility with the approach of the Torino model \cite{Rui16}.
Moreover getting the 2p2h responses written in terms of two 1p1h
response functions (or Lindhard functions) allows us to include a
phenomenological scaling function instead of the Lindhard function of
the free Fermi gas to evaluate the 2p2h responses. This opens the
possibility to extend the free Fermi gas 2p2h response functions to
what is expected from an interacting nucleus. In addition, this can be
also seen as an alternative way to include finite-size effects in the
Fermi gas 2p2h responses instead of the more common local density
approximation \cite{Ama94}.

\section{Acknowledgements}

This work has been partially supported by the Spanish Ministerio de
Economia y Competitividad and ERDF (European Regional Development
Fund) under contracts FIS2014-59386-P, FIS2014-53448-C2-1, by the Junta de
Andalucia (grants No. FQM-225, FQM160), by the INFN under project
MANYBODY, and part (TWD) by the U.S. Department of Energy under
cooperative agreement DE-FC02-94ER40818. IRS acknowledges support from
a Juan de la Cierva fellowship from MINECO (Spain). GDM acknowledges
support from a Junta de Andalucia fellowship (FQM7632, Proyectos de
Excelencia 2011).

\appendix

\section{Boundaries of the 1p1h responses}

In this appendix we derive the 
$\omega$-limits of the RFG response function, Eq. (\ref{r1p1h-scaling}).

We use the following result, which is easy to prove and states that if an
on-shell particle with momentum $p$ inside the Fermi gas takes
energy-momentum $(\omega,q)$, then the limit cases, $\np$ parallel or
anti-parallel to $\nq$, correspond to the condition
\begin{equation}
\left.
\begin{array}{c}
E_{p-q}=E+\omega \\
\mbox{or}\\
E_{p+q}=E+\omega \\
\end{array}
\right\}
\Leftrightarrow
\kappa \sqrt{1+1/\tau } - \lambda = \epsilon .
\label{eq:appA1}
\end{equation}
For fixed $q$, the limits of the quasielastic
 peak correspond to the scaling variable $\psi^2=1$;
see Eq. (\ref{r1p1h-scaling}).

From the definition of $\psi$, Eq. (\ref{psi2}), this corresponds  to
$\epsilon_0=\epsilon_F$, that is, a particle with Fermi momentum.
In the non-Pauli blocked regime, from Eq. (\ref{epsilon0}), the 
minimum energy for the particle is exactly $\epsilon$ as given in
Eq. (\ref{eq:appA1}). This implies that
\begin{equation}
\kappa \sqrt{1+1/\tau } - \lambda = \epsilon_F
\Leftrightarrow
\left\{
\begin{array}{c}
E_{k_F-q}=E_F+\omega \\
\mbox{or}\\
E_{k_F+q}=E_F+\omega \\
\end{array} .
\right.
\end{equation}
From this latter equation, by isolating $\omega$, one obtains the upper
and lower $\omega$-limits given in Eq. (\ref{omegalimits}).

\section{Cone-averaged propagator}

Here the integral appearing in the cone-averaged $\Delta$
propagator is computed
\begin{equation}
I(W,C)
\equiv
\int_0^{2\pi} 
 \frac{d\phi}{W-C\cos\phi} \, ,
\end{equation}
where $C>0$ and $W$ is a complex number. 
We transform the integral into a contour integral in the complex plane
on the variable $z=e^{i\phi}$. 
By multiplying and dividing by $z$ inside the integral we obtain
\begin{eqnarray}
I(W,C) 
&=&
 \int_0^{2\pi} 
\frac{ e^{i\phi}d\phi } {W e^{i\phi}-C \frac{e^{2i\phi}+1}{2}}
\nonumber\\
&=&
2i \oint \frac{dz}{Cz^2 -2Wz +C} ,
\end{eqnarray}
where the last integral is made along the unit circle counterclockwise.
The integral is evaluated by computing the poles inside the circle.
The poles are given by the roots of the second degree polynomial in the denominator, written in factorized form as
\begin{equation}
Cz^2 -2Wz +C = C(z-z_1)(z-z_2)
\end{equation}
where obviously $z_1z_2=1$ is satisfied, and
\begin{eqnarray}
z_1 &=& \frac{W}{C}+\frac{1}{C} \sqrt{W^2-C^2} \\
z_2 &=& \frac{W}{C}-\frac{1}{C} \sqrt{W^2-C^2} .
\end{eqnarray}
If $z_1$ is the pole inside the circle, then 
$z_2$ is automatically outside because $|z_1 z_2|=1$. 
Therefore there is only one pole inside the circle.
The integral is then computed as the residue at the pole
\begin{equation}
I(W,C) = \frac{2i}{C} 2\pi i 
\begin{array}[t]{c}\textstyle \rm Res \\ \scriptstyle |z_i|<1 \end{array}
\frac{1}{(z-z_1)(z-z_2)} .
\end{equation}
In the case $|z_1|<1$, this gives
\begin{equation}
I(W,C) = -\frac{4\pi}{C}\frac{1}{z_1-z_2} = -\frac{2\pi}{\sqrt{W^2-C^2}} .
\end{equation}
In the other case, $|z_1|> 1$, the pole inside the circle is  $z_2$, and the result is 
\begin{equation}
I(W,C) = -\frac{4\pi}{C}\frac{1}{z_2-z_1} = \frac{2\pi}{\sqrt{W^2-C^2}}.
\end{equation}


\begin{thebibliography}{99}


\bibitem{Don78}
  T.~W.~Donnelly, J.~W.~Van Orden, T.~De Forest, Jr. and W.~C.~Hermans,
  Phys.\ Lett.\  {\bf 76B} (1978) 393.


\bibitem{Van81}
  J.~W.~Van Orden and T.~W.~Donnelly,
  Ann. Phys.\  {\bf 131} (1981) 451.

\bibitem{Alb84}
  W.M. Alberico, M. Ericson, and A. Molinari, 
  Ann. Phys.\  {\bf 154} (1984) 356.

\bibitem{Mar09} M. Martini, M. Ericson, G. Chanfray, J. Marteau, 
                 Phys. Rev. C {\bf 80} (2009) 065501.

\bibitem{Mar10} M. Martini, M. Ericson, G. Chanfray, J. Marteau, 
Phys. Rev. C {\bf 81} (2010) 045502.

\bibitem{Nie11} J. Nieves, I. Ruiz Simo, M.J. Vicente Vacas, 
Phys. Rev. C {\bf 83} (2011) 045501.

\bibitem{Nie12} J. Nieves, I. Ruiz Simo, M.J. Vicente Vacas, 
Phys. Lett. B {\bf 707} (2012) 72.

\bibitem{Ama11} 
J.E. Amaro, M.B. Barbaro, J.A. Caballero, T.W. Donnelly, 
C.F. Williamson,
    Phys. Lett. B {\bf 696} (2011) 151.

\bibitem{Ama12} J.E. Amaro, M.B. Barbaro, J.A. Caballero, T.W. Donnelly, 
Phys. Rev. Lett. {\bf 108} (2012) 152501.

\bibitem{Gra13} R. Gran, J. Nieves, F. Sanchez, M.J. Vicente Vacas, 
Phys.Rev. D {\bf 88} (2013) 113007.

\bibitem{Gal16}
  K.~Gallmeister, U.~Mosel and J.~Weil,
  Phys.\ Rev.\ C {\bf 94} (2016)  035502.
 
\bibitem{Sob12} J.T. Sobczyk, Phys. Rev. C {\bf 86} (2012) 015504.


\bibitem{Alv17}
  L.~Alvarez-Ruso {\it et al.},
  arXiv:1706.03621 [hep-ph].


\bibitem{Mos16} U. Mosel, Ann. Rev. Nuc. Part. Sci. {\bf 66} (2016) 171.

\bibitem{Kat17}
  T.~Katori and M.~Martini,
  arXiv:1611.07770 [hep-ph].
  
\bibitem{Shn07} R. Shneor, {\em et al.,} (JLab Hall A Collaboration),
Phys. Rev. Lett. {\bf 99} (2007) 072501.

\bibitem{Sub08} R. Subedi, {\em et al.,}
Science {\bf 320} (2008) 1476.

\bibitem{Hen13}
  O.~Hen {\it et al.} [CLAS Collaboration],
  Phys.\ Lett.\ B {\bf 722} (2013) 63.

\bibitem{Hen14}
  O.~Hen {\it et al.} [CLAS Collaboration],
Science, {\bf 346} (2014) 614.

\bibitem{Ryc15} J. Ryckebusch, M. Vanhalst, W. Cosyn, 
J. Phys. G:
  Nucl. Part. Phys. {\bf 42} (2015) 055104.

\bibitem{Col15} C. Colle, O. Hen, W. Cosyn, I. Korover, E. Piasetzky,
  J. Ryckebusch, and L. B. Weinstein, Phys. Rev. C {\bf 92} (2015) 024604.

\bibitem{Col16} C. Colle, W. Cosyn, J. Ryckebusch,  
Phys. Rev. C {\bf 93} (2016) 034608.

\bibitem{Acc14} R. Acciarri, {\em et al.}, Phys. Rev. D {\bf 90} (2014) 012008.  

\bibitem{Wei16} L.B. Weinstein, O. Hen, E. Piasetzky,
Phys.Rev. C {\bf 94} (2016) 045501.

\bibitem{Nie16}  K. Niewczas, J. T. Sobczyk, Phys. Rev. C {\bf 93} (2016) 035502.

\bibitem{Van16} 
  T.~Van Cuyck, N.~Jachowicz, R.~Gonzalez-Jimenez, M.~Martini, V.~Pandey, J.~Ryckebusch and N.~Van Dessel,
  Phys.\ Rev.\ C {\bf 94} (2016) 024611.
  
  \bibitem{Van17}
  T.~Van Cuyck, N.~Jachowicz, R.~Gonzalez-Jimenez, J.~Ryckebusch and N.~Van Dessel,
  arXiv:1702.06402 [nucl-th].
  
\bibitem{Ama93}
  J.~E.~Amaro, G. Co', A.~M.~Lallena,
Ann. Phys.  {\bf 221} (1993) 306.

\bibitem{Ama94}
  J.~E.~Amaro, A.~M.~Lallena and G.~Co,
  Nucl.\ Phys.\ A {\bf 578} (1994) 365.

\bibitem{Alb91}
W.M. Alberico, A. De Pace, A. Drago, and A. Molinari,
Riv. Nuov. Cim. vol. {\bf14}, n.5 (1991) 1.

\bibitem{Gil97}
A. Gil, J. Nieves, and E. Oset,
Nucl. Phys. A {\bf 627} (1997) 543.

\bibitem{Dek91}
M.J. Dekker, P.J. Brussaard, and J.A. Tjon,
Phys. Lett. B {\bf 266} (1991) 249.

\bibitem{Dek92}
M.J. Dekker, P.J. Brussaard, and J.A. Tjon,
Phys. Lett. B {\bf 289} (1992) 255.

\bibitem{Dek94}
M.J. Dekker, P.J. Brussaard, and J.A. Tjon,
Phys. Rev. C {\bf 49} (1994) 2650.

\bibitem{DePace03}
  A.~De Pace, M.~Nardi, W.~M.~Alberico, T.~W.~Donnelly and A.~Molinari,
  Nucl.\ Phys.\ A {\bf 726} (2003) 303.

\bibitem{DePace04}
  A.~De Pace, M.~Nardi, W.~M.~Alberico, T.~W.~Donnelly and A.~Molinari,
  Nucl.\ Phys.\ A {\bf 741} (2004) 249.

\bibitem{Ama10}
  J.~E.~Amaro, C.~Maieron, M.~B.~Barbaro, J.~A.~Caballero and T.~W.~Donnelly,
  Phys.\ Rev.\ C {\bf 82} (2010) 044601.

\bibitem{Meg16a}
  G.~D.~Megias, J.~E.~Amaro, M.~B.~Barbaro, J.~A.~Caballero and T.~W.~Donnelly,
  Phys.\ Rev.\ D {\bf 94} (2016) 013012.

\bibitem{Rui16}
  I.~Ruiz Simo, J.~E.~Amaro, M.~B.~Barbaro, A.~De Pace, J.~A.~Caballero and T.~W.~Donnelly,
  J.Phys. G {\bf 44} (2017) 065105.
  
  \bibitem{Meg16b}
  G.~D.~Megias, J.~E.~Amaro, M.~B.~Barbaro, J.~A.~Caballero, T.~W.~Donnelly and I.~Ruiz Simo,
   Phys.\ Rev.\ D {\bf 94} (2016) 093004.
  
  \bibitem{Car92} R.C. Carrasco and E. Oset, Nucl. Phys. A {\bf 536} (1992) 445.

\bibitem{Rui17}
  I.~Ruiz Simo, J.~E.~Amaro, M.~B.~Barbaro, J.~A.~Caballero, G.~D.~Megias and T.~W.~Donnelly,
 Phys. Lett. B {\bf 770} (2017) 193.
  
  \bibitem{Ama05a} J.E. Amaro, M.B. Barbaro, J.A. Caballero, T.W. Donnelly
                 A. Molinari, and I. Sick,
                Phys. Rev. C {\bf 71} (2005) 015501.

\bibitem{Ama05b} J.E. Amaro, M.B. Barbaro, J.A. Caballero, T.W. Donnelly,
                 C. Maieron,                  
                Phys. Rev. C {\bf 71} (2005) 065501.

\bibitem{Ama95} J.E. Amaro, A.M. Lallena, G. Co',
Int. J. Mod. Phys. E {\bf 03} (1994) 735. 

\bibitem{Sim14}
  I.~Ruiz Simo, C.~Albertus, J.~E.~Amaro, M.~B.~Barbaro, J.~A.~Caballero and T.~W.~Donnelly,
  Phys.\ Rev.\ D {\bf 90} (2014) 033012.

\bibitem{Sim14b}
  I.~Ruiz Simo, C.~Albertus, J.~E.~Amaro, M.~B.~Barbaro, J.~A.~Caballero and T.~W.~Donnelly,
  Phys. Rev. D {\bf 90} (2014) 053010.

\bibitem{Nie04} J. Nieves, J.E. Amaro, and M. Valverde, 
Phys. Rev. C {\bf 70} (2004) 055503.

\bibitem{Hernandez:2007qq}
  E.~Hernandez, J.~Nieves and M.~Valverde,
  Phys.\ Rev.\ D {\bf 76} (2007) 033005.
  
\bibitem{Amaro:2017eah}
  J.~E.~Amaro, M.~B.~Barbaro, J.~A.~Caballero, A.~De Pace, T.~W.~Donnelly, G.~D.~Megias and I.~Ruiz Simo,
 Phys.Rev. C95 (2017) no.6, 065502. 
\end{thebibliography}
\end{document}